\pdfoutput=1

\documentclass[12pt]{article}

\usepackage{subeqnarray}
\usepackage{indentfirst} 
\usepackage{amsfonts,amssymb,amsmath,amsthm}
\usepackage{bm}          
\usepackage{graphicx}
\usepackage{cite}
\usepackage{url}
\usepackage[colorlinks=true,%
            citecolor=black,%
            linkcolor=black,%
            urlcolor=blue,%
            breaklinks=true]{hyperref}

\usepackage[normalem]{ulem}
\usepackage{xcolor}

\numberwithin{equation}{section}  

\begin{document}

\date{June 27, 2022 \\[1mm] revised September 12, 2022}

\title{Ergodicity of the Wang--Swendsen--Koteck\'y algorithm on
       several classes of lattices on the torus}

\author{
  {\small Jes\'us Salas} \\[-0.5mm]
  {\small\it Departamento de Matem\'aticas}  \\[-0.2cm]
  {\small\it Universidad Carlos III de Madrid} \\[-0.2cm]
  {\small\it Avda.\  de la Universidad, 30}    \\[-0.2cm]
  {\small\it 28911 Legan\'es}           \\[-2mm]
  {\small\it SPAIN}                     \\
  {\small\it Grupo de Teor\'{\i}as de Campos y F\'{\i}sica
             Estad\'{\i}stica}\\[-2mm]
  {\small\it Instituto Gregorio Mill\'an, Universidad Carlos III de
             Madrid}\\[-2mm]
  {\small\it Unidad Asociada al Instituto de Estructura de la Materia, CSIC}
             \\[-2mm]
  {\small\it SPAIN}           \\[-1mm]
  {\small\tt jsalas@math.uc3m.es} \\
  {\protect\makebox[5in]{\quad}}  
  \\[-3mm]
  {\small Alan D.~Sokal}                  \\[-0.5mm]
     {\small\it Department of Mathematics}   \\[-2mm]
     {\small\it University College London}   \\[-2mm]
     {\small\it Gower Street}                \\[-2mm]
     {\small\it London WC1E 6BT}             \\[-2mm]
     {\small\it UNITED KINGDOM}              \\[-2mm]
     {\small\tt sokal@math.ucl.ac.uk}        \\
     {\small\it Department of Physics}       \\[-2mm]
     {\small\it New York University}         \\[-2mm]
     {\small\it 726 Broadway}                \\[-2mm]
     {\small\it New York, NY 10003}          \\[-2mm]
     {\small\it USA}                         \\[-2mm]
     {\small\tt sokal@nyu.edu}               \\[3mm]
}

\bibliographystyle{plain}
\maketitle
\thispagestyle{empty}   

\begin{abstract}
We prove the ergodicity of the Wang--Swendsen--Koteck\'y (WSK) algorithm
for the zero-temperature $q$-state Potts antiferromagnet
on several classes of lattices on the torus.
In particular, the WSK algorithm is ergodic for $q\ge 4$ on any quadrangulation
of the torus of girth $\ge 4$.
It is also ergodic for $q \ge 5$ (resp.~$q \ge 3$)
on any Eulerian triangulation of the torus such that one sublattice 
consists of degree-4 vertices while the other two sublattices
induce a quadrangulation of girth $\ge 4$
(resp.~a bipartite quadrangulation) of the torus. 
These classes include many lattices of interest in statistical mechanics.

\end{abstract}


\clearpage

%
%
%
%
\newcommand{\be}{\begin{equation}}
\newcommand{\ee}{\end{equation}}
\newcommand{\<}{\langle}
\renewcommand{\>}{\rangle}
\newcommand{\widebar}{\overline}
\def\reff#1{(\protect\ref{#1})}
\def\spose#1{\hbox to 0pt{#1\hss}}
\def\ltapprox{\mathrel{\spose{\lower 3pt\hbox{$\mathchar"218$}}
 \raise 2.0pt\hbox{$\mathchar"13C$}}}
\def\gtapprox{\mathrel{\spose{\lower 3pt\hbox{$\mathchar"218$}}
 \raise 2.0pt\hbox{$\mathchar"13E$}}}
\def\textprime{${}^\prime$}
\def\half{\frac{1}{2}}
\def\third{\frac{1}{3}}
\def\twothird{\frac{2}{3}}
\def\smfrac#1#2{\textstyle \frac{#1}{#2}}
\def\smhalf{\smfrac{1}{2} }

%
%
\newcommand{\restrict}{\upharpoonright}
\newcommand{\drop}{\setminus}
\renewcommand{\emptyset}{\varnothing}
\newcommand{\eqdef}{\stackrel{\rm def}{=}}
\newcommand{\rem}{\textrm{rem}}
\newcommand{\Sym}{{\mathfrak{S}}}
\def\proof{\par\medskip\noindent{\sc Proof.\ }}
\def\sketchofproof{\par\medskip\noindent{\sc Sketch of proof.\ }}
\def\qed{ $\square$ \bigskip}
\newcommand{\myendremark}{ $\blacksquare$ \bigskip}
\def\proofof#1{\bigskip\noindent{\sc Proof of #1.\ }}
\newcommand{\myle}{\preceq}
\newcommand{\myge}{\succeq}
\newcommand{\mygt}{\succ}

%
%
\newcommand{\cyc}{{\rm cyc}}
\newcommand{\exc}{{\rm exc}}
\newcommand{\rec}{{\rm rec}}
\newcommand{\arec}{{\rm arec}}
\newcommand{\erec}{{\rm erec}}

%
%
\newcommand{\C}{{\mathbb C}}
\newcommand{\D}{{\mathbb D}}
\newcommand{\Z}{{\mathbb Z}}
\newcommand{\N}{{\mathbb N}}
\newcommand{\R}{{\mathbb R}}
\newcommand{\Q}{{\mathbb Q}}

%
%
\newcommand{\TT}{{\mathsf T}}
\newcommand{\HH}{{\mathsf H}}
\newcommand{\VV}{{\mathsf V}}
\newcommand{\JJ}{{\mathsf J}}
\newcommand{\PP}{{\mathsf P}}
\newcommand{\DD}{{\mathsf D}}
\newcommand{\QQ}{{\mathsf Q}}
\newcommand{\RR}{{\mathsf R}}

%
%
\newcommand{\bgamma}{{\bm{\gamma}}}
\newcommand{\bdelta}{{\bm{\delta}}}
\newcommand{\bmu}{{\bm{\mu}}}
\newcommand{\bsigma}{{\bm{\sigma}}}
\newcommand{\vecbsigma}{{\vec{\bm{\sigma}}}}
\newcommand{\bpi}{{\bm{\pi}}}
\newcommand{\vecbpi}{{\vec{\bm{\pi}}}}
\newcommand{\btau}{{\bm{\tau}}}
\newcommand{\bphi}{{\bm{\phi}}}
\newcommand{\bvarphi}{{\bm{\varphi}}}
\newcommand{\bGamma}{{\bm{\Gamma}}}

%
%
\newcommand{\psibar}{ {\bar{\psi}} }
\newcommand{\varphibar}{ {\bar{\varphi}} }

%
%
\newcommand{\bfa}{ {\bf a} }
\newcommand{\bfb}{ {\bf b} }
\newcommand{\bfc}{ {\bf c} }
\newcommand{\bfp}{ {\bf p} }
\newcommand{\bfr}{ {\bf r} }
\newcommand{\bfs}{ {\bf s} }
\newcommand{\bft}{ {\bf t} }
\newcommand{\bfu}{ {\bf u} }
\newcommand{\bfv}{ {\bf v} }
\newcommand{\bfw}{ {\bf w} }
\newcommand{\bfx}{ {\bf x} }
\newcommand{\bfy}{ {\bf y} }
\newcommand{\bfz}{ {\bf z} }
\newcommand{\bfT}{ {\bf T} }
\newcommand{\bone}{ {\mathbf 1} }
\newcommand{\bzero}{ {\mathbf 0} }

%
%
\newcommand{\ba}{{\bm{a}}}
\newcommand{\bb}{{\bm{b}}}
\newcommand{\bc}{{\bm{c}}}
\newcommand{\bd}{{\bm{d}}}
\newcommand{\bee}{{\bm{e}}}
\newcommand{\bff}{{\bm{f}}}
\newcommand{\bA}{{\bm{A}}}
\newcommand{\bB}{{\bm{B}}}
\newcommand{\bC}{{\bm{C}}}
\newcommand{\bP}{{\bm{P}}}
\newcommand{\bS}{{\bm{S}}}
\newcommand{\bT}{{\bm{T}}}

%
%
\newcommand{\scra}{{\mathcal{A}}}
\newcommand{\scrb}{{\mathcal{B}}}
\newcommand{\scrc}{{\mathcal{C}}}
\newcommand{\scrd}{{\mathcal{D}}}
\newcommand{\scre}{{\mathcal{E}}}
\newcommand{\scrf}{{\mathcal{F}}}
\newcommand{\scrg}{{\mathcal{G}}}
\newcommand{\scrh}{{\mathcal{H}}}
\newcommand{\scri}{{\mathcal{I}}}
\newcommand{\scrj}{{\mathcal{J}}}
\newcommand{\scrk}{{\mathcal{K}}}
\newcommand{\scrl}{{\mathcal{L}}}
\newcommand{\scrm}{{\mathcal{M}}}
\newcommand{\scrn}{{\mathcal{N}}}
\newcommand{\scro}{{\mathcal{O}}}
\newcommand{\scrp}{{\mathcal{P}}}
\newcommand{\scrq}{{\mathcal{Q}}}
\newcommand{\scrr}{{\mathcal{R}}}
\newcommand{\scrs}{{\mathcal{S}}}
\newcommand{\scrt}{{\mathcal{T}}}
\newcommand{\scru}{{\mathcal{U}}}
\newcommand{\scrv}{{\mathcal{V}}}
\newcommand{\scrw}{{\mathcal{W}}}
\newcommand{\scrx}{{\mathcal{X}}}
\newcommand{\scry}{{\mathcal{Y}}}
\newcommand{\scrz}{{\mathcal{Z}}}

%
%
\def\hboxrm#1{ {\hbox{\scriptsize\rm #1}} }
\def\hboxsans#1{ {\hbox{\scriptsize\sf #1}} }
\def\hboxscript#1{ {\hbox{\scriptsize\it #1}} }

%
%
\newtheorem{theorem}{Theorem}[section]
\newtheorem{definition}[theorem]{Definition}
\newtheorem{proposition}[theorem]{Proposition}
\newtheorem{lemma}[theorem]{Lemma}
\newtheorem{corollary}[theorem]{Corollary}
\newtheorem{conjecture}[theorem]{Conjecture}
\newtheorem{result}[theorem]{Result}
\newtheorem{question}[theorem]{Question}
\newtheorem{problem}[theorem]{Problem}

%
%
\newcommand{\stirlingsubset}[2]{\genfrac{\{}{\}}{0pt}{}{#1}{#2}}
\newcommand{\stirlingcycle}[2]{\genfrac{[}{]}{0pt}{}{#1}{#2}}
\newcommand{\associatedstirlingsubset}[2]%
      {\left\{\!\! \stirlingsubset{#1}{#2} \!\! \right\}}
\newcommand{\assocstirlingsubset}[3]%
      {{\genfrac{\{}{\}}{0pt}{}{#1}{#2}}_{\! \ge #3}}
\newcommand{\assocstirlingcycle}[3]{{\genfrac{[}{]}{0pt}{}{#1}{#2}}_{\ge #3}}
\newcommand{\associatedstirlingcycle}[2]{\left[\!\!%
            \stirlingcycle{#1}{#2} \!\! \right]}
\newcommand{\euler}[2]{\genfrac{\langle}{\rangle}{0pt}{}{#1}{#2}}
\newcommand{\eulergen}[3]{{\genfrac{\langle}{\rangle}{0pt}{}{#1}{#2}}_{\! #3}}
\newcommand{\eulersecond}[2]{\left\langle\!\! \euler{#1}{#2} \!\!\right\rangle}
\newcommand{\eulersecondBis}[2]{\big\langle\!\! \euler{#1}{#2} \!\!\big\rangle}
\newcommand{\eulersecondgen}[3]%
     {{\left\langle\!\! \euler{#1}{#2} \!\!\right\rangle}_{\! #3}}
\newcommand{\associatedstirlingcycleBis}[2]{\big[ \!\!%
            \stirlingcycle{#1}{#2} \!\! \big]}
\newcommand{\binomvert}[2]{\genfrac{\vert}{\vert}{0pt}{}{#1}{#2}}
\newcommand{\doublebinom}[2]{\left(\!\! \binom{#1}{#2} \!\!\right)}
\newcommand{\nueuler}[3]{{\genfrac{\langle}{\rangle}{0pt}{}{#1}{#2}}^{\! #3}}
\newcommand{\nueulergen}[4]%
{{\genfrac{\langle}{\rangle}{0pt}{}{#1}{#2}}^{\! #3}_{\! #4}}

%
%
\newenvironment{sarray}{
          \textfont0=\scriptfont0
          \scriptfont0=\scriptscriptfont0
          \textfont1=\scriptfont1
          \scriptfont1=\scriptscriptfont1
          \textfont2=\scriptfont2
          \scriptfont2=\scriptscriptfont2
          \textfont3=\scriptfont3
          \scriptfont3=\scriptscriptfont3
        \renewcommand{\arraystretch}{0.7}
        \begin{array}{l}}{\end{array}}

\newenvironment{scarray}{
          \textfont0=\scriptfont0
          \scriptfont0=\scriptscriptfont0
          \textfont1=\scriptfont1
          \scriptfont1=\scriptscriptfont1
          \textfont2=\scriptfont2
          \scriptfont2=\scriptscriptfont2
          \textfont3=\scriptfont3
          \scriptfont3=\scriptscriptfont3
        \renewcommand{\arraystretch}{0.7}
        \begin{array}{c}}{\end{array}}

\newcommand{\doi}[1]{\href{http://dx.doi.org/#1}{\texttt{doi:#1}}}
\newcommand{\arxiv}[1]{\href{http://arxiv.org/abs/#1}{\texttt{arXiv:#1}}}
\newcommand{\seqnum}[1]{\href{http://oeis.org/#1}{#1}}

\newcommand*{\Scale}[2][4]{\scalebox{#1}{$#2$}}%

 \newcommand\Kc{\kappa}  
 \newcommand\proofofcase[2]{\bigskip\noindent{\sc Case #1: #2.\ }}
 \newcommand\algcr{\mathop{algcr}}
%
%
\section{Introduction}

Proper $q$-colourings belong to the common ground for researchers in both 
statistical mechanics and combinatorics. 
For the former group, proper $q$-colourings appear naturally when considering
the $q$-state antiferromagnetic (AF) model Potts model on an (undirected) graph
\cite{Potts_52,Wu_82,Baxter_book,Wu_84,Sokal_bcc2005} 
in the limit of zero temperature ($T=0$). In this limit, 
the Gibbs measure is a uniform counting measure over 
the set of proper $q$-colourings. Usually, researchers use Monte Carlo (MC)
simulations \cite{Binder_Landau} to sample such a measure. The algorithm
of choice is usually the Wang--Swendsen--Koteck\'y (WSK) algorithm 
\cite{WSK1,WSK2}. This is a cluster Markov Chain MC, and it 
satisfies the required properties for its convergence to the
target probability distribution for any temperature $T > 0$.
However, antiferromagnets sometimes show interesting properties at $T=0$:
for example, some $q$-state Potts AF have a critical point at $T=0$ 
\cite{Kondev_96,Burton_Henley_97,SS_98,selfdual1,selfdual2}. 
Unfortunately, the ergodicity of the WSK algorithm is not guaranteed at $T=0$ 
(see e.g.\ \cite{Lubin_Sokal,MS1,MS2}); 
and the non-ergodicity of WSK at $T=0$ for some value of $q$ on 
a certain graph means that WSK may not converge to the desired target 
probability distribution. Moreover, the lack of ergodicity of the WSK 
algorithm at $T=0$ indicates that the autocorrelation time is likely to be 
very large for small temperatures $T > 0$,  
leading to undesirable artifacts at low temperature, 
as was observed for $q=4$ on the triangular lattice on the torus \cite{SS_tri}. 

For mathematicians, proper $q$-colourings on an undirected graph $G$ are 
counted by the chromatic polynomial $P_G(q)$. They define a ``dynamics'' based
on Kempe moves (see Section~\ref{sec.WSK}). 
Two $q$-colourings of $G$ that can be obtained by 
a finite number of Kempe moves are said to be $q$-equivalent
(or Kempe-equivalent). Moreover, 
$q$-equivalence is an equivalence relation on the set of proper $q$-colourings,
and the natural question is to know the number of such equivalence classes 
under Kempe moves. 

A key observation is that Kempe moves are the same as the WSK moves 
at $T=0$. Therefore, the existence of a unique Kempe equivalence class 
of $q$-colourings on a graph $G$ is equivalent to the ergodicity of the 
WSK algorithm for the zero-temperature $q$-state Potts AF on $G$. 
Therefore, results and ideas from combinatorialists can benefit 
researchers in statistical mechanics, and vice versa.

MC simulations are usually performed in periodic boundary
conditions, in order to avoid artifacts associated with the breaking
of translation invariance (i.e., surface effects).
In other words, one is studying a graph $G$ embedded on the torus.
But this sometimes
causes problems with ergodicity, which is more delicate for graphs on
the torus than for planar graphs.  For example, for the triangular
lattice of any size with {\em free}\/ boundary conditions (i.e.\ on the
plane), WSK is ergodic for $q=4$ (indeed, WSK-ergodicity for $q=4$
holds for any 3-colourable planar graph \cite{Mohar_07});  but for
$3m \times 3n$ triangular lattices with {\em periodic}\/ boundary
conditions (i.e.\ on the torus) --- which are indeed the natural
sizes, as they respect 3-colourability --- WSK is nonergodic for $q=4$ 
whenever $m,n \ge 2$ \cite{MS1}.

WSK is known to be ergodic for any $q\ge 2$ on any bipartite graph 
(see Theorem~\ref{thm.bipartite}),
which includes $2m\times 2n$ square and hexagonal
lattices with periodic boundary conditions. 
On the other hand, WSK is {\em not}\/ ergodic for $q=3$ on
periodic square lattices of size $3m\times 3n$ 
with $m,n$ relatively prime \cite{Lubin_Sokal}.  
Similarly, for triangular lattices on the torus, the ergodicity of WSK 
has been proven for $q\ge 7$ \cite{Jerrum_private,Mohar_07} 
(see Proposition~\ref{cor.JM}), $q=6$ \cite{Bonamy_19},
and $q=5$ \cite{Cranston_21};
but, as mentioned above, ergodicity can fail
for $q=4$ \cite{MS1}. Finally, for kagome lattices on the torus, the 
WSK-ergodicity has been proven for $q\ge 5$ \cite{Jerrum_private,Mohar_07} 
(see Proposition~\ref{cor.JM}) and $q=4$ \cite{McDonald,Bonamy_19};
but it can fail for $q=3$ \cite{MS2}.

These issues are relevant for physics because, as is well known,
universality does not hold generically for AF models, such as the
$q$-state AF Potts model:  rather, the phase diagram depends 
strongly on the microscopic details of the lattice where the model is defined. 
For this reason, MC simulations become a key tool for exploring the
phase diagram and critical behaviour of AF models.  
At this point, it is useful to introduce the temperature-like variable 
\be
v \;=\; e^{\beta J} -1
\label{def_v}
\ee
where $J$ is the coupling constant of the Potts model, and 
$\beta=(k_B\, T)^{-1}$; 
see Section~\ref{sec.Potts}. Therefore, the AF regime ($J<0$) corresponds to 
$v\in [-1,0)$.  

We expect the existence of an AF critical curve $v_\mathrm{AF}(q)$ 
that starts at the point $(q,v)=(0,0)$ and, as $q$ increases, the value of 
$v_\mathrm{AF}(q)$ decreases. (We extend the $q$-state Potts model to real
values of $q$ by using the Fortuin--Kasteleyn representation \cite{FK1,FK2}:
see Section~\ref{sec.Potts}.) 
This AF critical curve may be single-valued (as for the square lattice 
\cite{Baxter_book}) or have branches (as for the triangular lattice 
\cite{JSS}).
We define $q_c$ to be the largest value of $q$ for which $v_\mathrm{AF}(q_c)=-1$.
(Of course, $q_c$ depends on the lattice we are considering.) 
Now, if $q>q_c$, the system is always disordered for any $T\ge 0$; if 
$q=q_c$, the system has a critical point at $T=0$ ($v=-1$), and 
is disordered at any $T>0$ ($-1 < v < 0$);
and if $q<q_c$, any behaviour is possible. 
For the square (resp.\ triangular) lattice,
we have $q_c(\mathrm{Sq})=3$ (resp.\ $q_c(\mathrm{Tri})=4$). 

It was believed until fairly recently that the maximum value of $q_c$ for a 
plane quadrangulation (resp.~triangulation) is $q_c=3$ (resp.~$q_c=4$).
However, it is now known \cite{diced,planar_AF_largeq,selfdual1,selfdual2,BH} 
that this belief is false;
indeed, in this paper we will be particularly concerned with
2-dimensional (2D) lattices for which $q_c > 4$.
More specifically, we are mostly interested in quadrangulations
with $q_c \ge 4$, and triangulations with $q_c\ge 5$.
These classes of lattices are
the ones that are currently least understood
from the statistical-mechanics point of view. 

As a matter of fact, the value of $q_c$ for 2D lattices
is not even bounded in general. In 
Ref.~\cite{planar_AF_largeq}, there were introduced several 2D
graph families with arbitrary large values for $q_c$.
More precisely, in each of these families (say, $(F_n)_{n\ge 1}$),
each member $F_n$ is labeled with a positive integer $n$,
and $q_c(F_n)$ tends to infinity as $n \to\infty$.
All these families are decorations of square-lattice grids:
two of them ($G'_n$ and $G''_n$) are quadrangulations of the torus,
and one ($H'''_n$) is a triangulation of the torus;
see Section~\ref{subsec.def} below.
(In Ref.~\cite{diced}, another family with the same property was found, but 
its construction was more involved.)
Indeed, within these families we do find cases of physical interest. 
The numerical results of  Ref.~\cite{planar_AF_largeq}
showed that the phase transitions for $q\gtrsim 8$ were first-order,
while the order of the transitions for $4 < q \lesssim 8$ was inconclusive. 

\begin{figure}[t]
\centering
  \includegraphics[width=0.8\columnwidth]{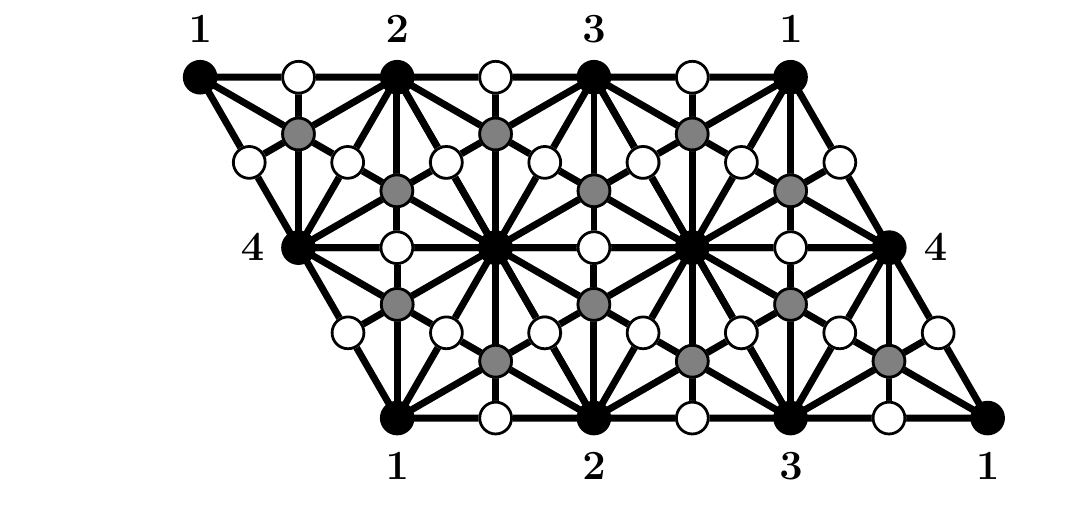} 
  \vspace*{-2mm}
  \caption{%
  Bisected-hexagonal lattice of size $3\times 2$ (unit cells) with periodic 
  boundary conditions. Vertices belonging to subsets $V$, $V'$, and $V_4$ are
  depicted as black, white, and gray dots, respectively. Points with the same
  label should be identified.
}
\label{fig_BH}
\end{figure}

Another type of interesting lattices were introduced in Ref.~\cite{union-jack}.
These ones are tripartite Eulerian triangulations of the torus such that one 
sublattice contains degree-4 vertices, while the other two induce a bipartite 
quadrangulation (see also \cite{selfdual1,selfdual2}). This class of
triangulations of the torus will be denoted $\mathcal{T}_0$ [see 
Definition~\ref{def_Ts}(a)], and the class of bipartite quadrangulations of the
torus $\mathcal{Q}_0$ [see Definition~\ref{def_Qs}(a)]. 
In Ref.~\cite{union-jack}, the $q$-state AF Potts model was studied on two
lattices of the class $\mathcal{T}_0$:
the bisected-hexagonal (BH) lattice (Fig.~\ref{fig_BH})
and the Union-jack (UJ) lattice (Fig.~\ref{fig_tri}). The former was
characterized by $q_c(\text{BH})=5.395(10)$, and the latter by 
$q_c(\text{UJ})=4.326(5)$. 

%
%


All the foregoing lattices share a common property: none of them is regular
(unlike the more common lattices such as
 square, triangular, hexagonal, and kagome \cite{SS_97}). 
This is probably why our knowledge of the phase diagram of the $q$-state 
AF Potts model on each of them is rather poor, and some of them might be hiding
some interesting phenomena. 

Using a field-theoretic approach, Delfino and Tartaglia \cite{Delfino_17} 
recently found, for 2D models in the continuum, an 
$\mathbb{S}_q$-invariant renormalization-group fixed point for 
$4\le q \le (7+\sqrt{17})/2 \approx 5.561553$, which could
potentially correspond to a critical $5$-state Potts AF on some 2D lattice.  
However, their approach could not predict which lattices $G$ were the 
``good'' ones, in the sense that the 5-state $G$-lattice AF Potts model 
would have a critical point.

Since $q_c(\text{BH})>5$ implies the existence of a phase transition at $q=5$,
the BH lattice emerged as a natural candidate to test
the ideas of Delfino and Tartaglia \cite{Delfino_17}.
The numerical results of Ref.~\cite{union-jack} initially suggested
hat the 5-state Potts AF on the BH lattice has
a critical point at $v=-0.95132(2)$
with critical exponents $X_m=0.113(4)$ and $X_t=0.495(5)$.
However, this model was reconsidered in Ref.~\cite{BH},
and the results supported the existence of a very weak 
first-order phase transition at $v_c=-0.951 308(2)$. Unfortunately, at that
time there was no result concerning the ergodicity of the WSK algorithm
for this particular case. This difficulty was overcome
empirically by a careful
analysis of the autocorrelation times \cite{Sokal_97} at low temperatures.
The numerical data suggested the conjecture
that the WSK algorithm is indeed ergodic at $T=0$ for this case. 
 
In the present paper, we prove the just-mentioned conjecture
concerning the ergodicity of the WSK algorithm for the $q=5$ 
model on the BH lattice, as a special case of much more general results. 
For triangulations in a class ${\mathcal T}_1 \supset {\mathcal T}_0$
(see Definition~\ref{def_Ts}), we show
[Theorem~\ref{theo.main.tri}(b)]
that WSK is ergodic at $T=0$ for $q \ge 5$;
and for triangulations in ${\mathcal T}_0$, we show
(Theorem~\ref{theo.main.tri2})
that it is ergodic for $q \ge 3$.\footnote{
   Here the case $q=3$ is trivial, as every triangulation
   in the class ${\mathcal T}_0$ is uniquely 3-colourable modulo permutations.
   So the interesting case is $q \ge 4$.
}
We also prove
[Corollary~\ref{cor.quad}(b)]
that WSK is ergodic for $q\ge 4$
on the class $\mathcal{Q}_1$ of quadrangulations of the torus
that have girth $\ge 4$;
this includes the class $\mathcal{Q}_0$ of bipartite quadrangulations
of the torus (but is strictly larger).
{}From a statistical-mechanical point of view, all the results of the present 
work ensure that MC simulations of the $q$-state AF Potts model can be safely 
performed at $T=0$ (and at small $T>0$)
on certain large classes of lattices on the torus. 
A second step, which goes beyond the scope of this paper,
will be to study the phase diagrams and
universality classes of these models at $q=4$ and $q=5$. 

The results of this paper are twofold. On the one hand, 
by invoking some rather deep graph-theoretic results
\cite{Fisk3,Vergnas_81,Feghali_17,Bonamy_19},
we obtain some simple corollaries concerning
the ergodicity of WSK on several classes of
graphs and lattices on the torus. 
On the other hand, in the 
particular case of $q=4$, we employ the algebraic-topology methods of 
Fisk \cite{Fisk1,Fisk2,Fisk3} to improve our results for triangulations 
$T\in \mathcal{T}_0$ for $q=4$.
Even though our methods are fairly simple, 
they lead to non-trivial results concerning the ergodicity of the WSK
algorithm for certain values of $q$ on certain graph families on the torus 
(see Sections~\ref{sec.application} and~\ref{sec.q=4}).

This paper is organized as follows. Section~\ref{sec.setup} contains the 
background material concerning the Potts model and graph theory,
and reviews some known results concerning the ergodicity of the WSK algorithm.
In Section~\ref{sec.torus} we prove {some sufficient conditions}
for the ergodicity of the WSK algorithm.
{In Section~\ref{sec.application} we apply these results}
to regular lattices on the torus: in particular,
to quadrangulations and triangulations.
In Section~\ref{sec.q=4} we study
the ergodicity of the WSK algorithm for 
$q=4$ on the class $\mathcal{T}_0$
of triangulations of the torus, using Fisk's methods.  
Finally, in Section~\ref{sec.summary} we summarize our findings.

%
%
\section{Preliminaries} \label{sec.setup}

In this section, we will discuss the background needed in this paper. 
In Section~\ref{sec.Potts} we summarize the main definitions
concerning the Potts model,
and in Section~\ref{sec.def} we summarize some needed definitions
from graph theory.
Section~\ref{sec.WSK} describes the WSK algorithm at zero temperature
and the Kempe moves, and then reviews some known results
concerning the ergodicity of the WSK dynamics.  

%
%
\subsection{Potts model} \label{sec.Potts}

Let $G = (V,E)$ be a finite undirected graph with vertex set $V$ and edge set
$E$. We can assume without loss of generality that $G$ is {\em simple}\/,
i.e.\ it has no self-loops or multiple edges; and we shall do so throughout 
this paper.\footnote{
   In the graph-theoretic literature, what we are here calling ``self-loops'' 
   --- that is, edges connecting a vertex to itself --- are called ``loops'' 
   {\em tout court}\/.
   We use the term ``self-loops'' in order to avoid any confusion
   vis-\`a-vis physicist readers, who sometimes (especially when discussing
   Feynman diagrams) use the term ``loop'' as a synonym for ``cycle''.
}
On each vertex $x\in V$, we place a spin $\sigma_x \in [q] = \{1,2,\ldots,q\}$
where $q\ge 1$ is an arbitrary integer. These spins interact with the 
Hamiltonian
\be
\mathcal{H}(\sigma) \;=\; 
-J \, \sum\limits_{\{i,j\}\in E} \delta_{\sigma_i,\sigma_j} 
\label{def_H} 
\ee 
where the sum is over all edges in $G$, $J$ is the coupling constant, and 
$\delta_{x,y}$ is the Kronecker delta. The \emph{partition function} of this 
model is given by 
\be
Z_G(q,\beta J) \;=\; \sum\limits_{ \sigma \colon V \to [q]} 
e^{-\beta \, \mathcal{H}(\sigma) } \;=\; 
\sum\limits_{ \sigma \colon V \to [q]} 
\exp \left( \beta J \, \sum\limits_{\{i,j\}\in E} \delta_{\sigma_i,\sigma_j} 
     \right) 
\label{def_ZG}
\ee 
where the outer sum is over all possible spin configurations, and 
$\beta=(k_B T)^{-1}$. Moreover,
$Z_G$ is actually a polynomial in $q$ and in the temperature-like variable  
$v=e^{\beta J} -1$;
namely, we have the {\em Fortuin--Kasteleyn representation}\/ \cite{FK1,FK2}
\be
Z_G(q, v) \;=\; \sum_{A \subseteq E}  q^{k(A)} \,  v^{|A|} \,,
\label{def_ZG_FK}
\ee
where the sum runs over all spanning subgraphs $(V,A)$ of $G$, and $k(A)$ is
the number of connected components of $(V,A)$. We can then
consider $q$ and $v$ as commuting indeterminates.  

The limit $\beta J\to -\infty$ corresponds to the zero-temperature limit of 
this model in the AF regime ($J<0$), and is equivalent to take $v=-1$ in
\eqref{def_ZG_FK}. In this limit, the partition function 
\eqref{def_ZG}/\eqref{def_ZG_FK} becomes
the chromatic polynomial $P_G(q)$ of $G$, and gives the number of proper
$q$-colourings of $G$. 

%
%
\subsection{Some graph-theoretic definitions}  \label{sec.def}

Let $G = (V,E)$ be a finite undirected (simple) graph with $|V|=n$ vertices 
and $|E|=m$ edges.
The {\em degree}\/ of a vertex is the number of its nearest neighbours.
The {\em maximum degree}\/ $\Delta(G)$,
{\em minimum degree}\/ $\delta(G)$,
and {\em average degree}\/ $d(G)$ of the graph $G$
are defined in the obvious way; note that $d(G) = 2m/n$.
The graph $G$ is called {\em $\Delta$-regular}\/
if every vertex has the same degree $\Delta$
(so that the maximum, minimum and average degrees all coincide).

The {\em chromatic number}\/ $\chi(G)$ is the smallest integer $q$
for which there exists a proper $q$-colouring.
Obviously, $\chi(G) \le \Delta(G) + 1$,
since for any $q \ge \Delta(G) + 1$
we can find a proper $q$-colouring by the ``greedy algorithm'':
just take the vertices in any order, and successively colour them
in any allowable way;
since each vertex has at most $\Delta(G)$ nearest neighbours,
there is always an unused colour available.
Furthermore, Brooks' theorem \cite[Section~V.1, Theorem~3]{Bollobas}
says that if $G$ is a connected graph, then $\chi(G) \le \Delta(G)$
except when $G$ is a complete graph or an odd cycle.

But these bounds can be strengthened further.
The {\em degeneracy number}\/ $D(G)$ is defined as
$D(G) = \max\limits_{H \subseteq G} \delta(H)$,
where the max runs over all subgraphs
(or equivalently, all induced subgraphs) $H \subseteq G$.
If $D(G) \le d$, the graph $G$ is said to be {\em $d$-degenerate}\/.
Since $\delta(H) \le \Delta(H) \le \Delta(G)$,
we obviously have $D(G) \le \Delta(G)$.
Moreover, if $G$ is connected and not $\Delta$-regular,
then $D(G) \le \Delta(G) - 1$.
[{\sc Proof:}
We have $\delta(G) < \Delta(G)$ since $G$ is not $\Delta$-regular;
and every subgraph $H \subsetneq G$ has at least one vertex
adjacent to a vertex of $G \setminus H$
(since $G$ is connected),
so this vertex has degree $< \Delta(G)$ in $H$, making $\delta(H) < \Delta(G)$.
Hence $\delta(H) \le \Delta(G) - 1$ for all $H \subseteq G$.]
However, $D(G)$ can sometimes be {\em vastly}\/ smaller than $\Delta(G)$;
we will give some examples later. 
Here we only mention the graph $G_n$ defined in Section~\ref{sec.application} 
[see Fig.~\ref{fig_Gns}(a)], which has $\Delta(G_n)=4n$ but $D(G_n)=2$. 

An equivalent definition of degeneracy number is the following \cite{Lick}:
For a given ordering $v_1,\ldots,v_n$ of the vertex set $V$,
let $d_i$ be the degree of the vertex $v_i$
in the graph $G \setminus \{v_1,\ldots,v_{i-1}\}$,
and then define $d^\star = \max(d_1,\ldots,d_n)$;
then the degeneracy number $D(G)$ is the minimum of the values $d^\star$
over all orderings of $V$.
That is, we successively remove the vertices $v_1,\ldots,v_n$
from the graph, and $d_i$ is the degree of the vertex $v_i$
at the time of its removal;
then $d^\star$ is the largest degree encountered in this process;
and $D(G)$ is the smallest value of $d^\star$ that can be obtained
by a suitably chosen ordering of $V$.
In fact, a suitable ordering can always be found by the
{\em greedy algorithm}\/:  that is, take $v_i$ to be any vertex
of minimal degree in the graph
$G \setminus \{v_1,\ldots,v_{i-1}\}$.
We record these equivalent formulations as follows:

\begin{lemma}[Conditions for a graph to be $d$-degenerate]
   \label{lemma.degenerate}
For a graph $G$ and a positive integer $d$,
the following conditions are equivalent:
\begin{itemize}
   \item[(a)]  $G$ is $d$-degenerate.
   \item[(b)]  There exists an ordering $v_1,\ldots,v_n$ of the vertices of $G$
      such that, if $d_i$ is the degree of $v_i$ in the graph
      $G \setminus \{v_1,\ldots,v_{i-1}\}$,
      then $d_i \le d$ for all $i$ ($1 \le i \le n$).
   \item[(c)]  Let $v_1,\ldots,v_n$ be any ordering of the vertices of $G$
      in which $v_i$ is a vertex of minimal degree in the graph
      $G \setminus \{v_1,\ldots,v_{i-1}\}$;
      and let $d_i$ be that degree.
      Then $d_i \le d$ for all $i$ ($1 \le i \le n$).
\end{itemize}
\end{lemma}

It is now clear that $\chi(G) \le D(G) + 1$,
since for any $q \ge D(G) + 1$
we can find a proper $q$-colouring by employing the ``greedy algorithm''
in the order $v_n,\ldots,v_1$
--- that is, colour each vertex with any colour that has not been used
by its already-coloured neighbours ---
since each vertex has at most $D(G)$ neighbours
at the time it is inserted and coloured.

The degeneracy number will play an important role in our considerations,
as will be seen shortly.

Finally, the {\em girth}\/ of a graph $G$ is the length of the
shortest cycle in $G$;
if $G$ has no cycles (i.e.\ is a forest), then the girth is $\infty$.
Note that the girth is always $\ge 3$,
since our graphs have neither self-loops (i.e.\ cycles of length~1)
nor multiple edges (i.e.\ cycles of length~2).
A graph is called {\em triangle-free}\/ if it does not contain
any cycles of length 3, or in other words if its girth is $\ge 4$.

%
%
\subsection{WSK and Kempe moves: Known results} \label{sec.WSK}

The WSK algorithm \cite{WSK1,WSK2} reduces at zero temperature to
the following procedure: We choose at random two distinct colours 
$a,b\in [q]$, and let $G_{ab}$ be the induced subgraph of $G$ 
consisting of vertices $x\in V$ for which $\sigma_x\in \{a,b\}$. 
This induced subgraph is in general disconnected. Then, independently for each 
connected component of $G_{ab}$, we either interchange
the colours $a$ and $b$ on it, or leave the component unchanged;
each possibility is chosen with probability $1/2$.
This algorithm leaves invariant the uniform measure over the set of 
proper $q$-colourings of $G$, but its ergodicity cannot be taken for granted. 

A Kempe move consists in one basic WSK move at $T=0$: that is, we choose
two distinct colours, and we swap colours in one of the connected 
components of the induced subgraph $G_{ab}$. 
Two proper $q$-colourings $c_1$ and $c_2$ of $G$ are
said to be {\em $q$-equivalent}\/
(or {\em Kempe-equivalent}\/)
if one can be obtained from the other by means of a finite sequence of 
Kempe moves. 
It is obvious that $q$-equivalence is an equivalence relation on the set of
proper $q$-colourings; each equivalence class is called a {\em Kempe class}\/.
Clearly, the WSK algorithm is ergodic for $q$-colourings of $G$
$\Longleftrightarrow$
all proper $q$-colourings of $G$ are $q$-equivalent
$\Longleftrightarrow$
there is a unique Kempe class for $q$-colourings of $G$.
(Let us remark that if the graph $G$ is not $q$-colourable,
 then these properties hold vacuously, as the set of proper $q$-colourings
 of $G$ is empty.  But this trivial case is obviously of no interest!)

Let us now review some known sufficient conditions for the ergodicity
of the WSK algorithm.

One key tool in proving WSK-ergodicity is the following lemma,
due to Las Vergnas and Meyniel \cite[Lemma~2.3]{Vergnas_81}:

\begin{lemma}[Las Vergnas--Meyniel \cite{Vergnas_81}]
   \label{lemma.vergnas}
Let $G$ be a graph, and let $w$ be a vertex of $G$ of degree $< q$.
If all $q$-colourings of the graph $G \setminus w$ are $q$-equivalent,
then also all $q$-colourings of $G$ are $q$-equivalent.
\end{lemma}

\noindent
Please observe that this result is non-vacuous:
if the graph $G \setminus w$ is $q$-colourable,
then so is $G$, since the ``new'' vertex $w$ has fewer than $q$ neighbours.

So we can add new vertices of degree $< q$
and not only maintain the colourability,
but also maintain the ergodicity of the WSK algorithm.
In particular, we can apply this fact repeatedly,
starting from the empty graph,
building up the graph $G$ by adding vertices one at a time.
Recalling the second definition of degeneracy number,
we obtain the immediate corollary
\cite[Proposition~2.1]{Vergnas_81},
\cite[Proposition~2.4]{Mohar_07}
\cite[second part of Lemma~1]{Cranston_21}:

\begin{theorem}[Las Vergnas--Meyniel \cite{Vergnas_81}]
   \label{thm.vergnas}
If the graph $G$ is $d$-degenerate, then for all $q \ge d+1$,
all $q$-colourings of $G$ are $q$-equivalent.
\end{theorem}

\noindent
Once again this result is non-vacuous, because $\chi(G) \le D(G)+1$.
Indeed, Theorem~\ref{thm.vergnas} can be regarded as a
WSK-style strengthening of the result $\chi(G) \le D(G)+1$:
namely, for $q \ge D(G) + 1$,
the set of proper $q$-colourings is nonempty
{\em and}\/ consists of a single Kempe class.

Applying the bounds $D(G) \le \Delta(G)$,
and $D(G) \le \Delta(G)-1$ whenever $G$ is connected
and not $\Delta$-regular, we obtain the following weakened
version of Theorem~\ref{thm.vergnas}:

\begin{corollary}[Jerrum \cite{Jerrum_private}, Mohar \cite{Mohar_07}]
\label{cor.JM}
Let $G$ be a graph of maximum degree $\Delta$.  Then:
\begin{itemize}
   \item[(a)] For all $q \ge \Delta+1$,
       all $q$-colourings of $G$ are $q$-equivalent.
   \item[(b)] If, in addition, $G$ is connected and not $\Delta$-regular,
       then all $\Delta$-colourings of $G$ are $\Delta$-equivalent.
\end{itemize}
\end{corollary}

\begin{figure}[t]
\centering
  \includegraphics[width=0.2\columnwidth]{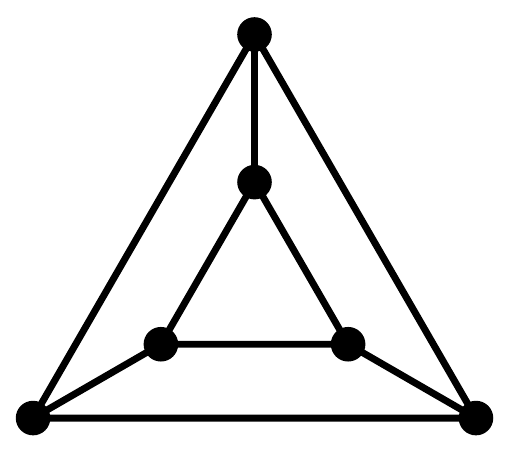}
  \vspace*{-2mm}
  \caption{%
  The triangular prism.  
}
\label{fig_3prism}
\end{figure}

In fact, the requirement in part~(b) that $G$ not be $\Delta$-regular
can be removed, by virtue of the following recent result:

\begin{theorem}[Feghali \emph{et al.} \cite{Feghali_17},
                Bonamy \emph{et al.} \cite{Bonamy_19}] 
   \label{thm.feghali-bonamy}
Let $G$ be a $\Delta$-regular graph with $\Delta\ge 3$;
and if $\Delta=3$, assume that $G$ is not the triangular prism
(Fig.~\ref{fig_3prism}).
Then all $\Delta$-colourings of $G$ are $\Delta$-equivalent.
\end{theorem} 

\noindent
Let us remark that if $G$ is the complete graph $K_{\Delta+1}$, then this 
result holds vacuously because $G$ has no proper $\Delta$-colouring.
In all other cases, however, Brooks' theorem guarantees that
$G$ does have at least one proper $\Delta$-colouring,
and the result is non-vacuous.
 
%
%

Another result concerns bipartite graphs and was independently obtained by
Burton--Henley \cite{Burton_Henley_97}, Ferreira--Sokal 
\cite{Ferreira_Sokal_99}, and  Mohar \cite{Mohar_07}:   

\begin{theorem} \label{thm.bipartite} 
Let $G$ be a bipartite graph. Then, for all $q \ge 2$,
all $q$-colourings of $G$ are $q$-equivalent. 
\end{theorem} 

Please note that all the foregoing results hold for arbitrary graphs,
which need not be embedded in a surface.
On the other hand, for planar graphs
there is a further sufficient condition for WSK-ergodicity,
in terms of the chromatic number $\chi(G)$:

\begin{theorem}[Mohar {\cite[Corollary~4.5]{Mohar_07}}]
   \label{thm.mohar}
Let $G$ be a planar graph.
Then, for all $q > \chi(G)$, all $q$-colourings of $G$ are $q$-equivalent.
\end{theorem}

In particular, we have:

\begin{theorem}[Meyniel {\cite{Meyniel}}]
   \label{thm.meyniel}
Let $G$ be a planar graph.
Then all 5-colourings of $G$ are 5-equivalent.
\end{theorem}

Theorem~\ref{thm.meyniel} is of course an immediate consequence of
Theorem~\ref{thm.mohar} together with the 4-colour theorem;
but Meyniel's \cite{Meyniel} proof does not require the 4-colour theorem.

Finally, a related result was proven recently by
Feghali \cite{Feghali_21}.
Recall that a graph is called {\em $q$-critical}\/
if it is $q$-colourable but not $(q-1)$-colourable,
but every proper subgraph is $(q-1)$-colourable.
Then:

\begin{theorem}[Feghali {\cite{Feghali_21}}]
   \label{thm.feghali}
Let $G$ be a 4-critical planar graph.
Then all 4-colourings of $G$ are 4-equivalent.
\end{theorem}
 
\begin{figure}[tbh]
\vspace*{-1cm}
\centering
  \begin{tabular}{c@{\hspace*{2cm}}c} 
  \includegraphics[width=0.3\columnwidth]{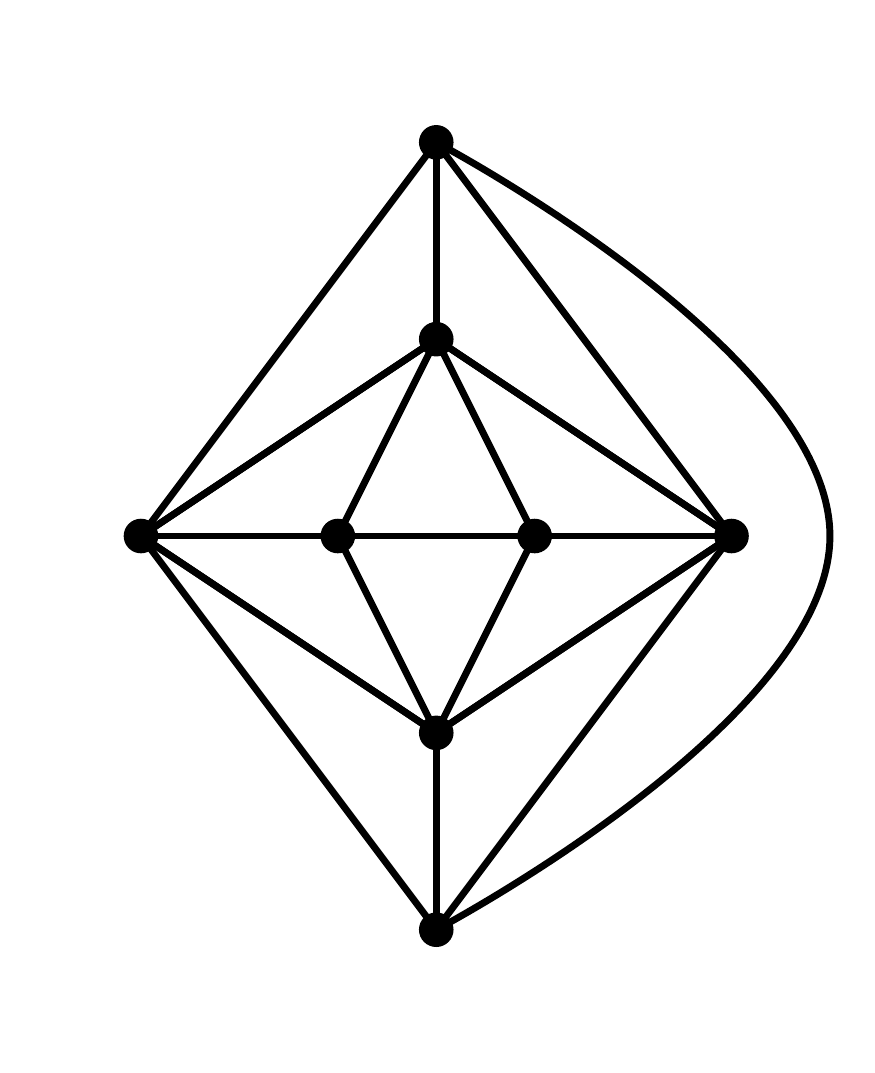} &  
  \includegraphics[width=0.3\columnwidth]{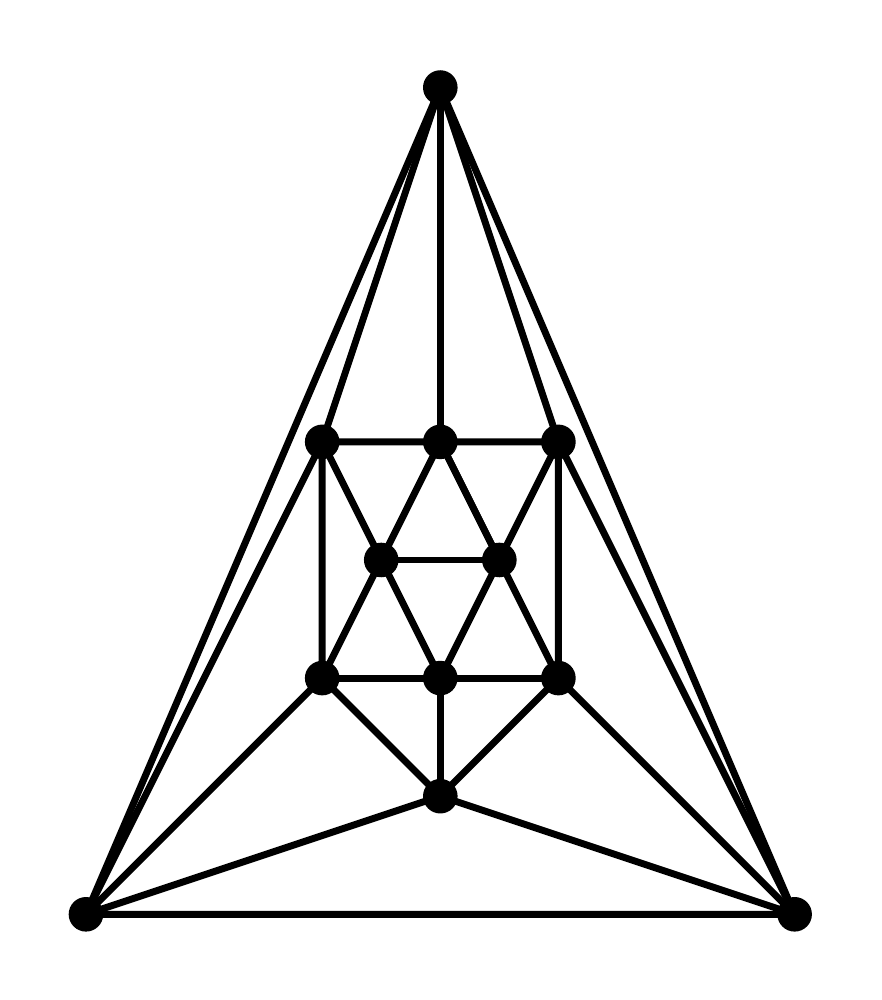} \\
  (a) & (b) \\[4mm] 
  \end{tabular} 
  \vspace*{-2mm}
  \caption{%
   (a) The graph $F_1$.
   (b) The icosahedron.
}
\label{fig_graphs2-5}
\end{figure}

\medskip
\noindent
{\bf Remarks.} 
   1.  When $\chi(G) = 2$ (i.e., $G$ is bipartite),
the restriction of Theorem~\ref{thm.mohar} to $q > \chi(G)$ is suboptimal
and is improved by Theorem~\ref{thm.bipartite} to $q \ge \chi(G)$.
However, when $\chi(G) = 3$ or 4, the bound $q > \chi(G)$
of Theorem~\ref{thm.mohar} is indeed best possible. 
For $\chi(G) = 3$, one obvious example is the 3-prism (see 
Fig.~\ref{fig_3prism}):
it has $\chi=3$, but there are two Kempe classes for 3-colourings.
For $\chi(G) = 4$, one example is the graph $F_1$ \cite[Fig.~1${}'$]{Meyniel},
shown here in Fig.~\ref{fig_graphs2-5}(a): it is planar and has $\chi(F_1)=4$,
but there are two Kempe classes for 4-colourings.
(One Kempe class consists of a single configuration modulo permutations,
while the other Kempe class consists of two configurations
modulo permutations.)
In the same vein, we have the icosahedron \cite[Fig.~8A]{Fisk2},
shown here in Fig.~\ref{fig_graphs2-5}(b):
again $\chi=4$, but there are 10 Kempe equivalence classes for $q=4$.

%
%

2. We can also find toroidal graphs $G$ such that there is no WSK-ergodicity
for $q = \chi(G)+1$. There are infinitely many examples described in 
\cite{MS1}:  namely, the triangular grids of size
$3m\times 3n$ with $m,n\ge 2$.
Here $\chi=3$, but there are at least two Kempe equivalence classes for
4-colourings. 

%
%
\section{Graphs on the torus} \label{sec.torus}

A (connected closed) {\em surface}\/ is, by definition,
a connected compact Hausdorff topological space $S$
that is locally homeomorphic to $\R^2$.
We recall the classification of surfaces
\cite[Section~3.1]{Mohar-Thomassen}:
Every surface is homeomorphic to precisely one
of the orientable surfaces $S_0, S_1, S_2, \ldots$
(where $S_p$ is the sphere with $p$ handles)
or one of the non-orientable surfaces $N_1, N_2, \ldots$
(where $N_q$ is the sphere with $q$ crosscaps).
In the former case, the {\em Euler characteristic}\/ is $\chi(S_p) = 2-2p$;
in the latter case, it is $\chi(N_q) = 2-q$.
In particular, the sphere $S_0$ has Euler characteristic~2,
and the torus $S_1$ has Euler characteristic~0.

A graph $G$ is {\em embedded}\/ in a surface $S$
if it is drawn so that the edges are simple paths
that intersect only at the endpoints.
It is a {\em cellular embedding}\/ if, in addition,
each face (i.e.\ each connected component of $S \setminus G$)
is homeomorphic to an open disc.\footnote{
   By \cite[Theorems~3.2.4 and 3.3.1]{Mohar-Thomassen},
   this is equivalent to the concept of a
   {\em 2-cell embedding}\/ as defined on
   \cite[p.~78]{Mohar-Thomassen}
   and employed in \cite[Section~3.1]{Mohar-Thomassen}.
}
Note that a cellularly embedded graph is necessarily connected
\cite[p.~78]{Mohar-Thomassen},
since otherwise one of the connected components of $S \setminus G$
would fail to be simply connected.
Henceforth we restrict attention to cellular embeddings.
{\em Euler's formula}\/ \cite[Lemma~3.1.4]{Mohar-Thomassen}
states that for a (connected) graph $G$ with $n$ vertices, $m$ edges and $f$ 
faces that is cellularly embedded
in a surface of Euler characteristic $\chi$,
we have
\be
   n \,-\, m \,+\, f  \;=\;  \chi
   \;.
\label{euler.fom}
\ee
In this paper, we will focus on graphs
that are cellularly embedded in the sphere ($\chi=2$)
or the torus ($\chi=0$);
in particular, all such graphs are connected.

One simple consequence of Euler's formula is the following well-known result: 

\begin{proposition}[Average degree of planar and toroidal graphs]
   \label{prop.euler.1}
\hfill\break
\vspace*{-6mm}
\begin{itemize}
   \item[(a)]
  Every planar graph of girth $g \: (\ge 3)$
       has average degree $< 2g/(g-2)$.
   \item[(b)]
  Every toroidal graph of girth $g \: (\ge 3)$
       has average degree $\le 2g/(g-2)$.
\end{itemize}
\end{proposition}

\proof
Assume first that $G$ is bridgeless, so that each edge lies on
the boundary of exactly two faces.
If $f_i$ is the number of faces of size $i$, then
\be
   2 m \;=\; \sum\limits_i i\, f_i \;\ge\; g f \,.
\ee
This implies that  $f \le (2/g) m$,
and if we plug this inequality into Euler's formula \eqref{euler.fom}, we get
\be
   \chi + m \;=\; n + f \;\le\; n + \frac{2}{g}\, m
   \quad \Longrightarrow \quad
   m \;\le\; \frac{g}{g-2} (n - \chi)  \,.
 \label{eq.proof.prop.euler.1}
\ee
Therefore, the average degree $d(G)$ satisfies
\be
   d(G) \;=\; \frac{2m}{n}
   \;\le\;
   \frac{2g}{g-2} \Bigl( 1 - \frac{\chi}{n} \Bigr)  \,.
\ee
Now use $\chi=2$ for planar graphs and $\chi=0$ for toroidal graphs.

For the general case, we prove the inequality \reff{eq.proof.prop.euler.1}
by induction on the number of bridges in $G$.
If $G$ has a bridge $e$, then the graph $G/e$ obtained by contracting $e$
has $n-1$ vertices and $m-1$ edges and one less bridge than $G$,
so by the inductive hypothesis we have
\be
   m-1 \;\le\; \frac{g}{g-2} (n - 1 - \chi)  \,.
\ee
This implies
\be
   m \;\le\; \frac{g}{g-2} (n - \chi)
\ee
since $g > 2$.
\qed 

Proposition~\ref{prop.euler.1} has the following
easy corollary concerning degeneracy:

\begin{corollary}[Degeneracy of planar and toroidal graphs]
   \label{cor.euler.2}
\hfill\break
\vspace*{-6mm}
\begin{itemize}
   \item[(a)] Every planar graph is 5-degenerate.
   \item[(b)] Every triangle-free planar graph is 3-degenerate.
   \item[(c)] Every planar graph of girth $\ge 6$ is 2-degenerate.
   \item[(d)] Every toroidal graph is 6-degenerate.
   \item[(e)] Every triangle-free toroidal graph is 4-degenerate.
   \item[(f)] Every toroidal graph of girth $\ge 5$ is 3-degenerate.
   \item[(g)] Every toroidal graph of girth $\ge 7$ is 2-degenerate.
\end{itemize}
\end{corollary}

\proof
(a) Every subgraph $H \subseteq G$ is planar;
by Proposition~\ref{prop.euler.1}(a) it has average degree $< 6$
and hence has a vertex of degree $\le 5$.

(b) Every subgraph $H \subseteq G$ is planar and has girth $\ge 4$;
by Proposition~\ref{prop.euler.1}(a) it has average degree $< 4$
and hence has a vertex of degree $\le 3$.

(c) Every subgraph $H \subseteq G$ is planar and has girth $\ge 6$;
by Proposition~\ref{prop.euler.1}(a) it has average degree $< 3$
and hence has a vertex of degree $\le 2$.

(d) Every subgraph $H \subseteq G$ is toroidal;
by Proposition~\ref{prop.euler.1}(b) it has average degree $\le 6$
and hence has a vertex of degree $\le 6$.

(e) Every subgraph $H \subseteq G$ is toroidal and has girth $\ge 4$;
by Proposition~\ref{prop.euler.1}(b) it has average degree $\le 4$
and hence has a vertex of degree $\le 4$.

(f) Every subgraph $H \subseteq G$ is toroidal and has girth $\ge 5$;
by Proposition~\ref{prop.euler.1}(b) it has average degree $\le 10/3 < 4$
and hence has a vertex of degree $\le 3$.

(g) Every subgraph $H \subseteq G$ is toroidal and has girth $\ge 7$;
by Proposition~\ref{prop.euler.1}(b) it has average degree $\le 14/5 < 3$
and hence has a vertex of degree $\le 2$.
\qed

\medskip

{\bf Remark.}
Result (a) is well known and can be found in \cite{Lick};
results (b), (c), and (d) are mentioned without proof in 
Refs.~\cite{Fihavc_02}, \cite{Hetherington_12}, and \cite{Cranston_21}, 
respectively.
\myendremark

\medskip

By combining Corollary~\ref{cor.euler.2} and Theorem~\ref{thm.vergnas},
we can deduce the following sufficient conditions for WSK-ergodicity: 

\begin{corollary}[Ergodicity of WSK for graphs on the plane and the torus]
   \label{cor.euler.3}
\hfill\break
\vspace*{-6mm}
\begin{itemize}
   \item[(a)] For every planar graph, WSK is ergodic for $q \ge 6$.
   \item[(b)] For every triangle-free planar graph,
       WSK is ergodic for $q \ge 4$.
   \item[(c)] For every planar graph of girth $\ge 6$,
       WSK is ergodic for $q \ge 3$. 
   \item[(d)] For every toroidal graph, WSK is ergodic for $q \ge 7$. 
   \item[(e)] For every triangle-free toroidal graph,
       WSK is ergodic for $q \ge 5$.
   \item[(f)] For every toroidal graph of girth $\ge 5$,
       WSK is ergodic for $q \ge 4$.
   \item[(g)] For every toroidal graph of girth $\ge 7$,
       WSK is ergodic for $q \ge 3$.
\end{itemize}
\end{corollary}

Here (a) is weaker than Meyniel's \cite{Meyniel} result
(Theorem~\ref{thm.meyniel} above)
that WSK is ergodic for $q \ge 5$ on every planar graph.
Furthermore,
(b) can alternatively be derived as
a consequence of Theorem~\ref{thm.mohar},
since it is known that every triangle-free planar graph
is 3-colorable \cite{Groetzsch,Thomassen_94,Thomassen_03}.
Also, (c) can be improved to $q \ge 2$:
if the graph is bipartite, then Theorem~\ref{thm.bipartite}
proves ergodicity for $q=2$;
and if the graph is not bipartite,
then there are no 2-colourings and the result holds vacuously.
Finally, (d) is mentioned without proof in \cite{Cranston_21}. 

But (d,e,f,g) can all be improved, as follows:

\begin{theorem}[Ergodicity of WSK for graphs on the torus]
   \label{thm.main}
\hfill\break
\vspace*{-7mm}
\begin{itemize}
   \item[(d$\,{}'$)] For every toroidal graph, WSK is ergodic for $q \ge 6$
        \cite{Cranston_21}.
   \item[(e$\,{}'$)] For every triangle-free toroidal graph,
       WSK is ergodic for $q \ge 4$.
   \item[(g$\:{}'$)] For every toroidal graph of girth $\ge 6$,
       WSK is ergodic for $q \ge 3$.
\end{itemize}
\end{theorem}

\proof
(d${}'$) Let $G$ be a toroidal graph;
the proof is by induction on the number of vertices in $G$.
By Proposition~\ref{prop.euler.1}(b),
the graph $G$ has average degree $\le 6$.
If $G$ is 6-regular, then the claim follows from
Theorem~\ref{thm.feghali-bonamy}.
If $G$ is not 6-regular, then $G$ has a vertex $w$ of degree $\le 5$.
By the inductive hypothesis, WSK is ergodic on $G \setminus w$ for $q \ge 6$.
Then Lemma~\ref{lemma.vergnas} implies that the same holds true for $G$.

(e${}'$, g$\,{}'$)  The proofs are analogous:
here the key fact is that the girth of $G \setminus w$
is at least as large as that of $G$.
And in case~(g$\,{}'$), note that the graph $G$ 
cannot be the triangular prism,
because the 3-prism has girth 3.
\qed

{\bf Remarks.} 1. Analogously to (c),
the result (g${}'$) can be improved to $q \ge 2$:
if the graph is bipartite, then Theorem~\ref{thm.bipartite}
proves ergodicity for $q=2$;
and if the graph is not bipartite,
then there are no 2-colourings and the result holds vacuously.

2. On the other hand,
(e${}'$) is best possible: there exist triangle-free toroidal graphs
such that WSK is not ergodic on them for $q=3$.
One example is any
square-lattice strip graph with periodic boundary conditions and size
$3m \times 3n$ with relatively prime integers $m,n > 1$ \cite{Lubin_Sokal}.
\myendremark

\medskip

It is, as far as we know, an open question whether
Corollary~\ref{cor.euler.3}(b) can be improved to ergodicity also for $q=3$.
Let us state this formally:

\begin{question}
   \label{question.trifree.q=3}
Is it true that for every triangle-free planar graph,
WSK is ergodic for $q=3$?
\end{question}

By Corollary~\ref{cor.JM} and Theorems~\ref{thm.feghali-bonamy}
and \ref{thm.bipartite}, any counterexample would have to be
planar, triangle-free, non-bipartite, and have maximum degree $\ge 4$.
We tested a few simple such graphs,
such as the $5_{\rm P} \times 3_{\rm F}$ and $7_{\rm P} \times 3_{\rm F}$
square lattices and graphs obtained from them by adding edges on the
top and/or bottom faces;
but all of the graphs we tried had a single Kempe class.

\bigskip

Another open problem (to the best of our knowledge)
is whether Theorem~\ref{thm.main}(d${}'$)
can be improved to ergodicity also for $q=5$.
We again state this formally:

\begin{question}
   \label{question.toroidal.q=5}
Is it true that for every toroidal graph, WSK is ergodic for $q=5$?
\end{question}

We tested several 6-regular triangulations $G$ of the torus
that are not 4-colourable
(see the list in \cite{Cranston_21}),
on the theory that these might be the most likely to yield counterexamples.
All of the graphs we tried had a single 5-Kempe class,
or there was no 5-colouring at all.
So we were unable to find any counterexample.

%
%
\section{Applications to regular lattices on the torus} \label{sec.application}

\subsection{Definitions of some families of lattices}  \label{subsec.def}

\begin{figure}[tbh]
\centering
  \includegraphics[width=0.6\columnwidth]{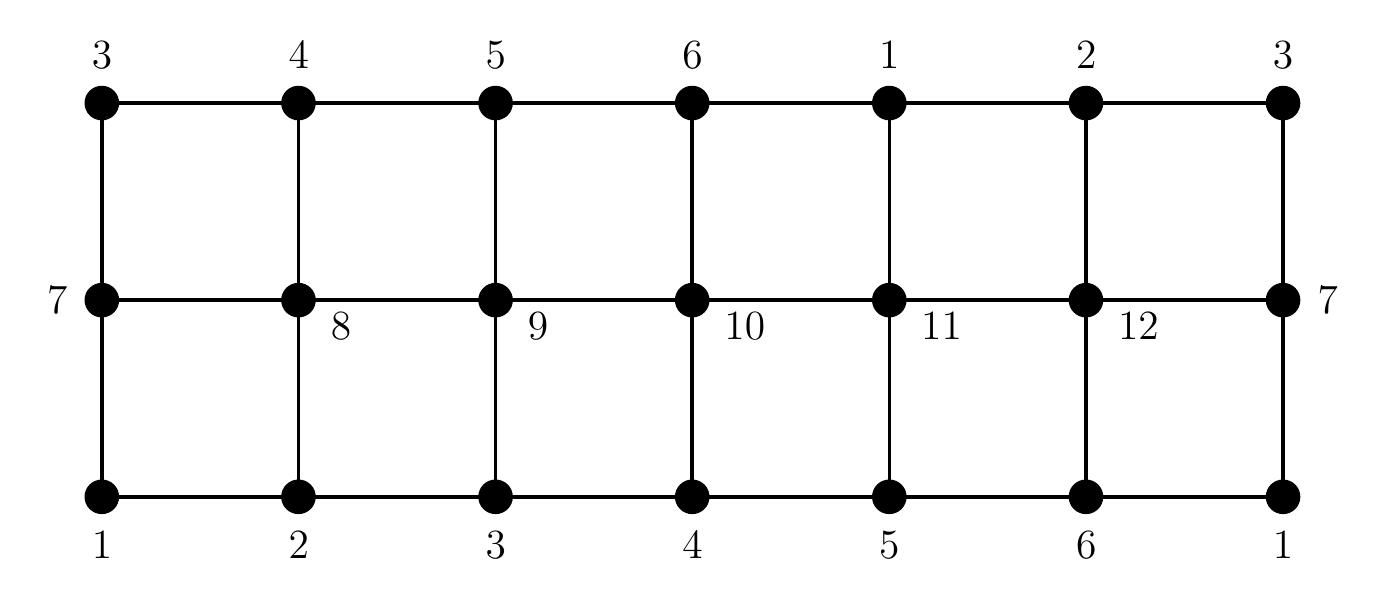}  
  \vspace*{-2mm}
  \caption{%
   Square lattice $\mathrm{Sq}(6,2,2)$ of size $6\times 2$ with periodic 
   boundary conditions in the horizontal direction, but the top and bottom
   rows are identified after cyclically shifting the top row by $2$ edges  
   to the right. 
}
\label{fig_sq_6x2}
\end{figure}

In Ref.~\cite{planar_AF_largeq}, several infinite families of graphs based on 
the square grid were introduced. To simplify the notation, let us denote
by $\mathrm{Sq}(a,b,c)$ an $a\times b$ square grid with 
``twisted periodic'' boundary conditions of twist $c$. That is, we start 
from a square grid of size $(a+1) \times (b+1)$, so that the top and bottom 
rows have $a$ edges, and the left and right sides have $b$ edges; we then embed
this graph on the torus by identifying the left and right sides
as usual, but identifying the top and bottom rows after
cyclically shifting the top row by $c$ edges to the right. 
Fig.~\ref{fig_sq_6x2} shows the bipartite quadrangulation 
$\mathrm{Sq}(6,2,2)$.
When $c=0$, we simply write $\mathrm{Sq}(a,b)=\mathrm{Sq}(a,b,0)$. 
In this case we assume $a,b \ge 3$, in order to avoid creating multiple edges.

%
%

\begin{figure}[t]
\centering
  \begin{tabular}{cccc}
  \includegraphics[width=0.22\columnwidth]{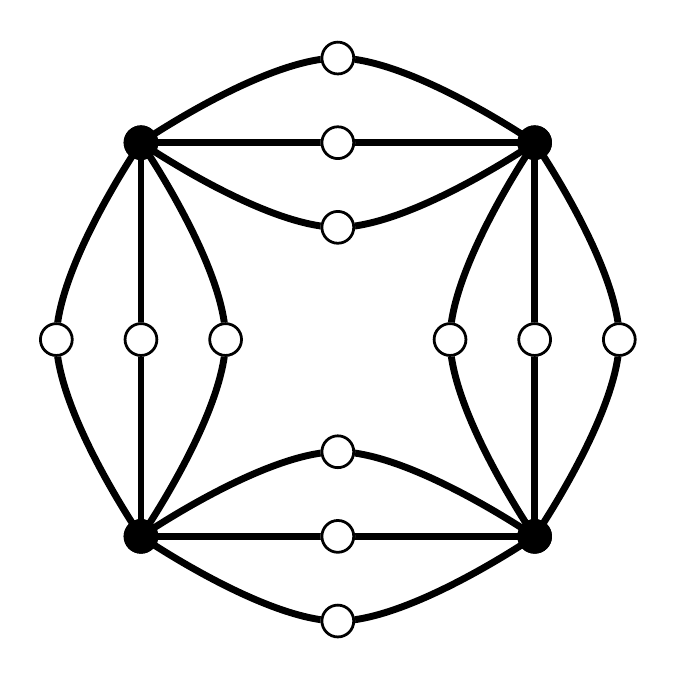} & 
  \includegraphics[width=0.22\columnwidth]{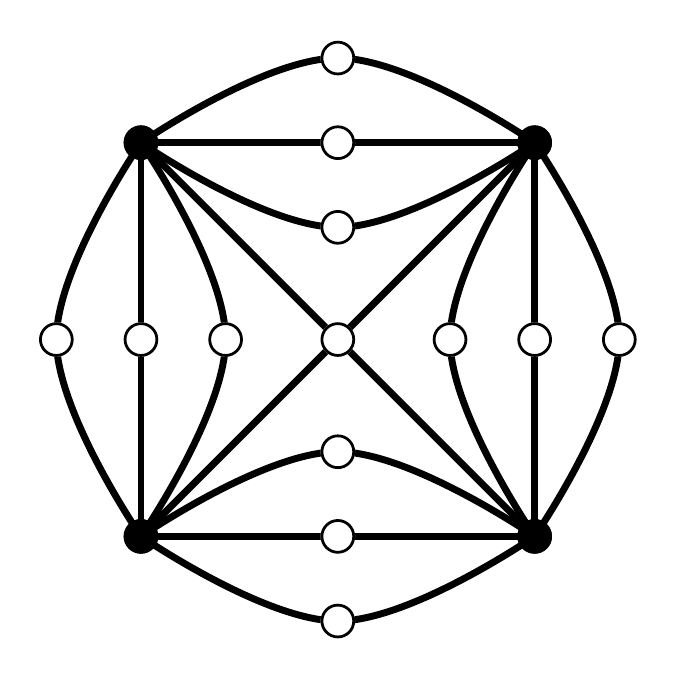} & 
  \includegraphics[width=0.22\columnwidth]{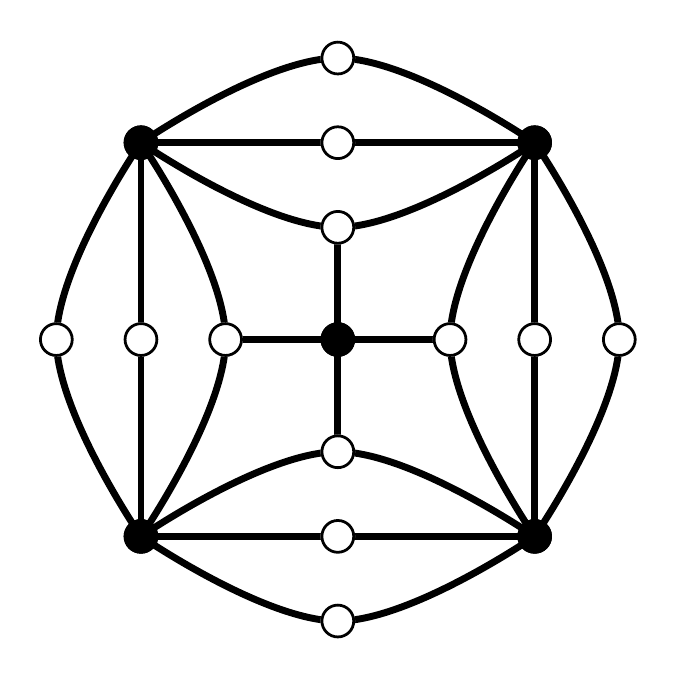} & 
  \includegraphics[width=0.22\columnwidth]{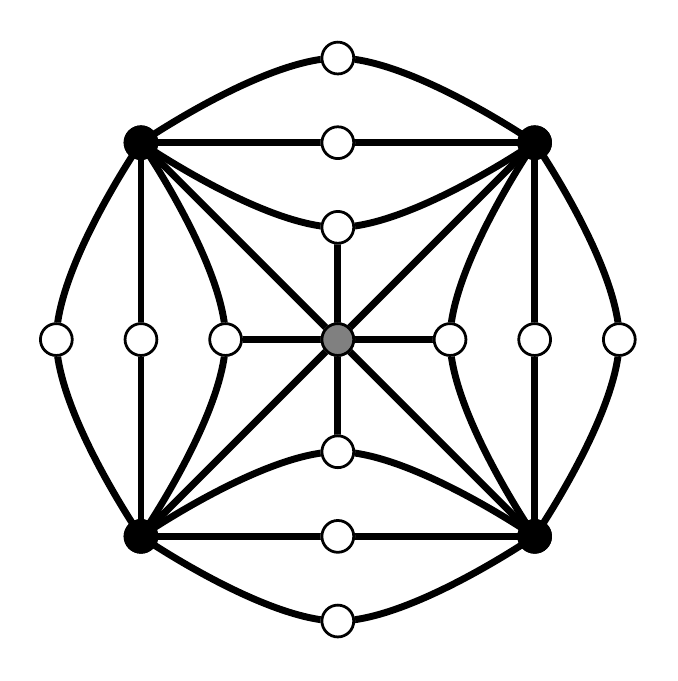} \\
  (a)   & (b) & (c) & (d) \\ 
  \end{tabular}
  \vspace*{-2mm}
  \caption{%
  Unit cells for the lattices $G_n$ (a), $G_n'$ (b), $G_n''$ (c),
  and $G_n'''$ (d) for $n=3$. Vertices coloured alike belong to the 
  same sublattice. 
}
\label{fig_Gns}
\end{figure}

\begin{figure}[t]
\centering
  \begin{tabular}{cccc}
  \includegraphics[width=0.22\columnwidth]{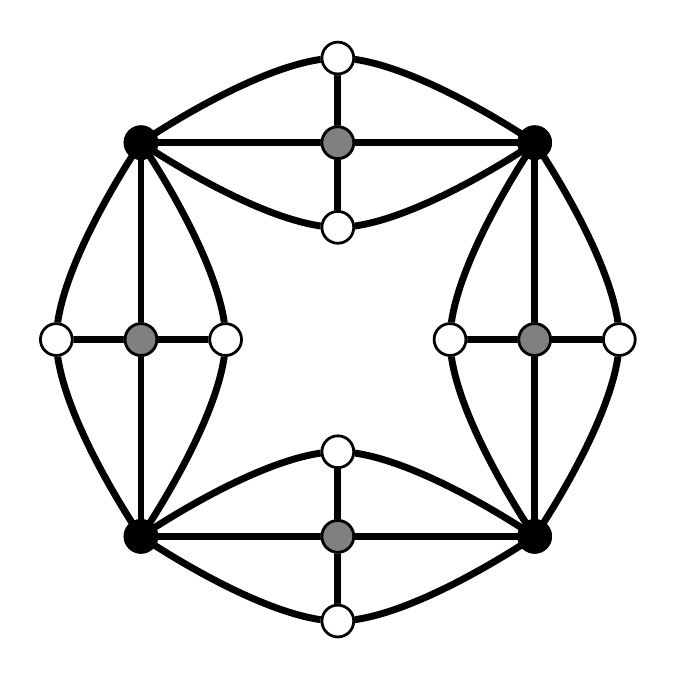} & 
  \includegraphics[width=0.22\columnwidth]{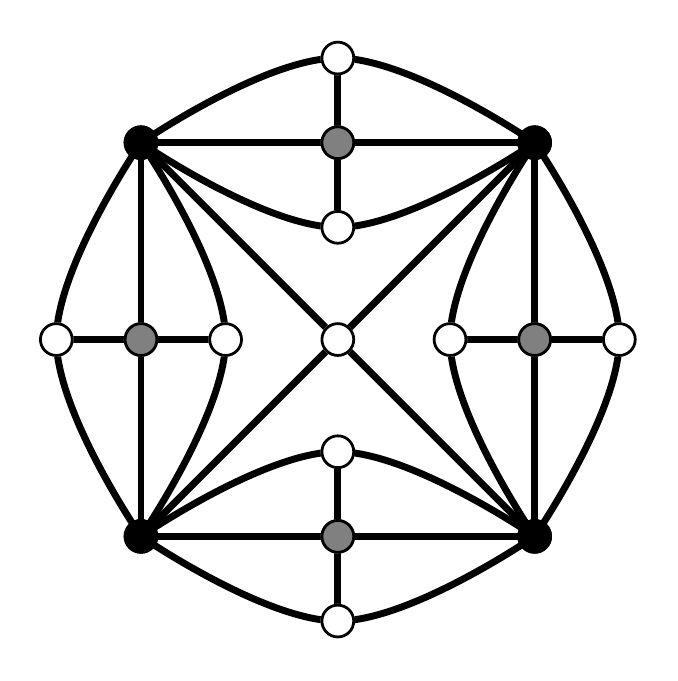} & 
  \includegraphics[width=0.22\columnwidth]{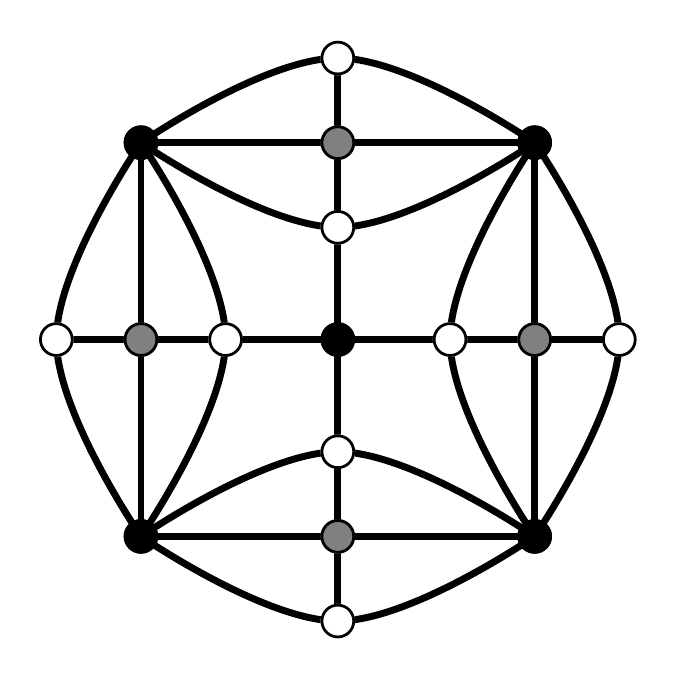} & 
  \includegraphics[width=0.22\columnwidth]{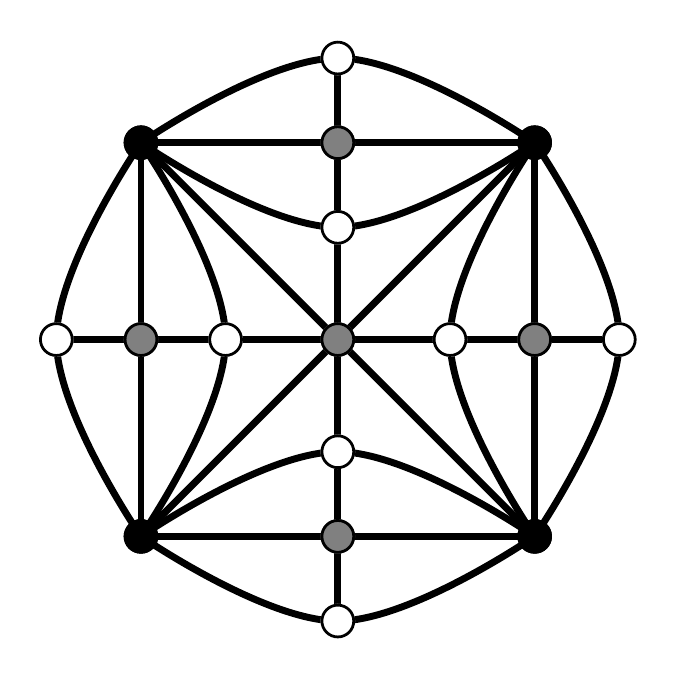} \\
  (a)   & (b) & (c) & (d) \\ 
  \end{tabular}
  \vspace*{-2mm}
  \caption{%
  Unit cells for the lattices $H_n$ (a), $H_n'$ (b), $H_n''$ (c),
  and $H_n'''$ (d) for $n=3$. Vertices coloured alike belong to the 
  same sublattice. 
}
\label{fig_Hns}
\end{figure}

For $a,b \ge 3$, we let
$G_n(a,b)$ [see Fig.~\ref{fig_Gns}(a)]
be the graph obtained from 
$\mathrm{Sq}(a,b)$ by replacing each edge by $n$ two-edge paths in parallel. 
Likewise, $H_n(a,b)$ [see Fig.~\ref{fig_Hns}(a)] is obtained from $G_n(a,b)$ 
by connecting each group of $n$ ``new parallel'' vertices with an 
$(n-1)$-edge path. Both $G_n(a,b)$ and $H_n(a,b)$ are neither triangulations  
nor quadrangulations of the torus.
Both $G_n(a,b)$ and $H_n(a,b)$ have maximum degree $4n$.
$G_n(a,b)$ is 2-degenerate;
it is bipartite whenever $a$ and $b$ are even;
and it has girth~4 whenever $a,b \ge 4$.\footnote{ 
   Notice that the girth of $G_n(a,b)$ is $g=3$ whenever $a$ and/or $b$ 
   are equal to $3$. 
}
Likewise, $H_n(a,b)$ is 3-degenerate;
it is tripartite (i.e.~has chromatic number~3)
whenever $a$ and $b$ are even;
and it has girth~3 (recall that $a,b \ge 3$).

%
%

%
%

We can obtain new modified families from both $G_n(a,b)$ and $H_n(a,b)$. 
The families $G'_n(a,b)$ [see Fig.~\ref{fig_Gns}(b)] and $H'_n(a,b)$ 
[see Fig.~\ref{fig_Hns}(b)] 
are obtained by adding a new vertex inside
each octagonal face of $G_n(a,b)$ 
or $H_n(a,b)$, respectively, and connecting it to the four surrounding 
vertices of the original square lattice $\mathrm{Sq}(a,b)$. 
Now $G'_n(a,b)$ is a 4-degenerate quadrangulation of the torus;
it is bipartite whenever $a$ and $b$ are even;
and it has girth~4 whenever $a,b \ge 4$.
$H'_n(a,b)$ is a 4-degenerate toroidal graph;
it is tripartite whenever $a$ and $b$ are even;
and it has girth~3.

The families $G_n''(a,b)$ [see Fig.~\ref{fig_Gns}(c)] and $H''_n(a,b)$ 
[see Fig.~\ref{fig_Hns}(c)],
by contrast, are obtained by adding a new vertex inside each octagonal face
of $G_n(a,b)$ or $H_n(a,b)$, respectively, and connecting
it to the four ``new'' surrounding vertices.
$G_n''(a,b)$ is a 3-degenerate quadrangulation of the torus;
it is bipartite whenever $a$ and $b$ are even;
and it has girth~4 whenever $a,b \ge 4$.
$H''_n(a,b)$ is a 4-degenerate toroidal graph;
it is tripartite whenever $a$ and $b$ are even;
and it has girth~3.

Finally, the families $G_n'''(a,b)$ [see Fig.~\ref{fig_Gns}(d)] and 
$H'''_n(a,b)$ [see Fig.~\ref{fig_Hns}(d)] are obtained by adding a new vertex 
inside each octagonal face of $G_n(a,b)$ or $H_n(a,b)$, respectively, and 
connecting it to {\em both}\/
the four surrounding vertices of the original square lattice 
$\mathrm{Sq}(a,b)$ {\em and}\/
to the four ``new'' surrounding vertices. 
$G_n'''(a,b)$ is a 4-degenerate toroidal graph;
it is tripartite whenever $a$ and $b$ are even;
and it has girth~3.
$H'''_n(a,b)$ is a 4-degenerate Eulerian triangulation of the torus;
it is tripartite whenever $a$ and $b$ are even;
and it has girth~3. Note also that when $n$ is odd,
the white sublattice of $H'''_n(a,b)$ contains only degree-4 vertices.

All these families of graphs have great interest for statistical mechanics
because $q_c$ tends to infinity with $n$ for all of them: in particular,
for $F_n=G_n,H_n,G'_n,H_n',G''_n,H'''_n$, 
$q_c(F_n)$ grows asymptotically like
$2n/W(2n) \sim 2n/\log n$ \cite{planar_AF_largeq} 
where $W$ is the Lambert $W$ function
defined by $W(x) e^{W(x)} = x$ \cite{Corless_96}.\footnote{
   This is probably also true for $H''_n$ and $G'''_n$,
   but $q_c$ for these lattices was not considered in \cite{planar_AF_largeq}.
}
Notice that for all these graphs, the maximum degree is $\Delta \ge 4n$,
while the degeneracy number is between $2$ and $4$.
In Ref.~\cite{diced}, another family with arbitrary large values of $q_c$ 
was found, but its construction is more involved. 

In addition to these lattices, we will consider in Section~\ref{sec.quad}
a class of quadrangulations of the torus investigated in 
\cite{selfdual1,selfdual2}. 
Finally, two classes of triangulations (obtained from such quadrangulations) 
will be presented in Section~\ref{sec.tri}. These triangulations of the 
torus have been studied in \cite{union-jack,BH}.
 
\subsection{Quadrangulations of the torus} \label{sec.quad}

\begin{figure}[t]
\centering
  \begin{tabular}{cc}
  \includegraphics[width=0.44\columnwidth]{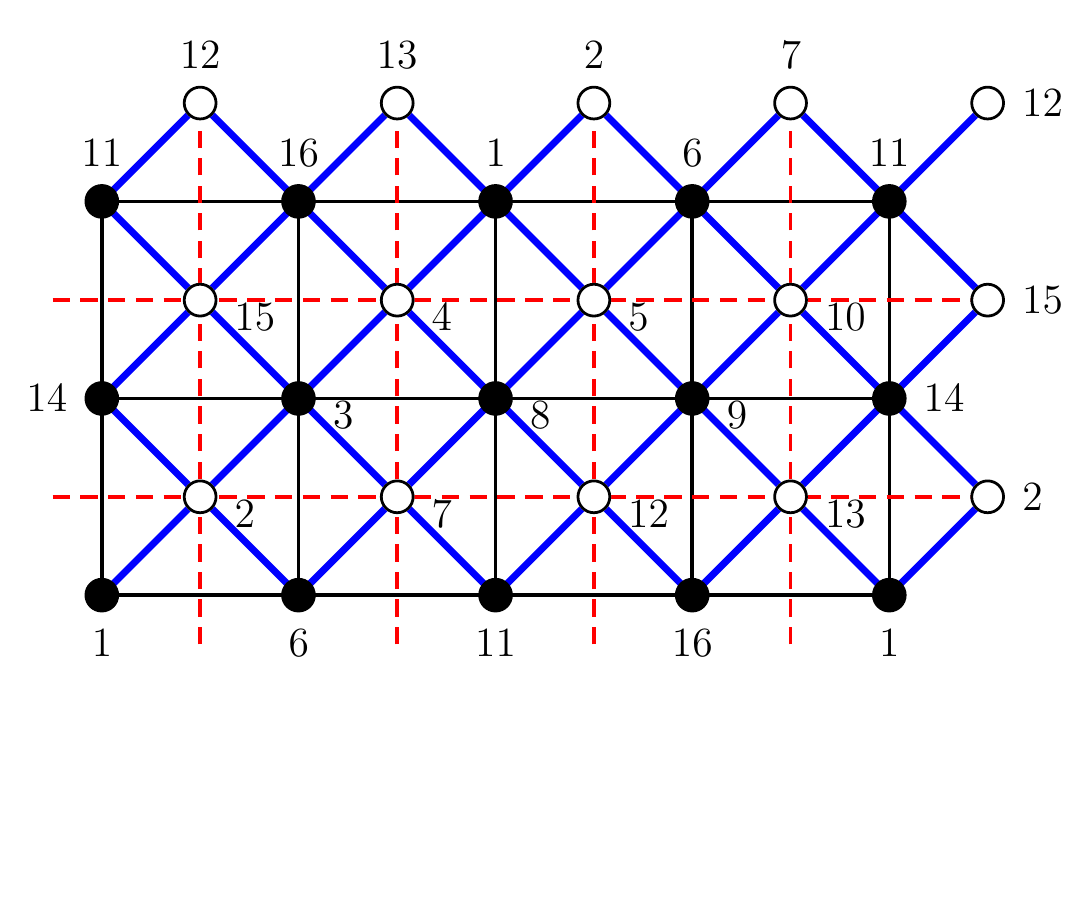} & 
  \includegraphics[width=0.44\columnwidth]{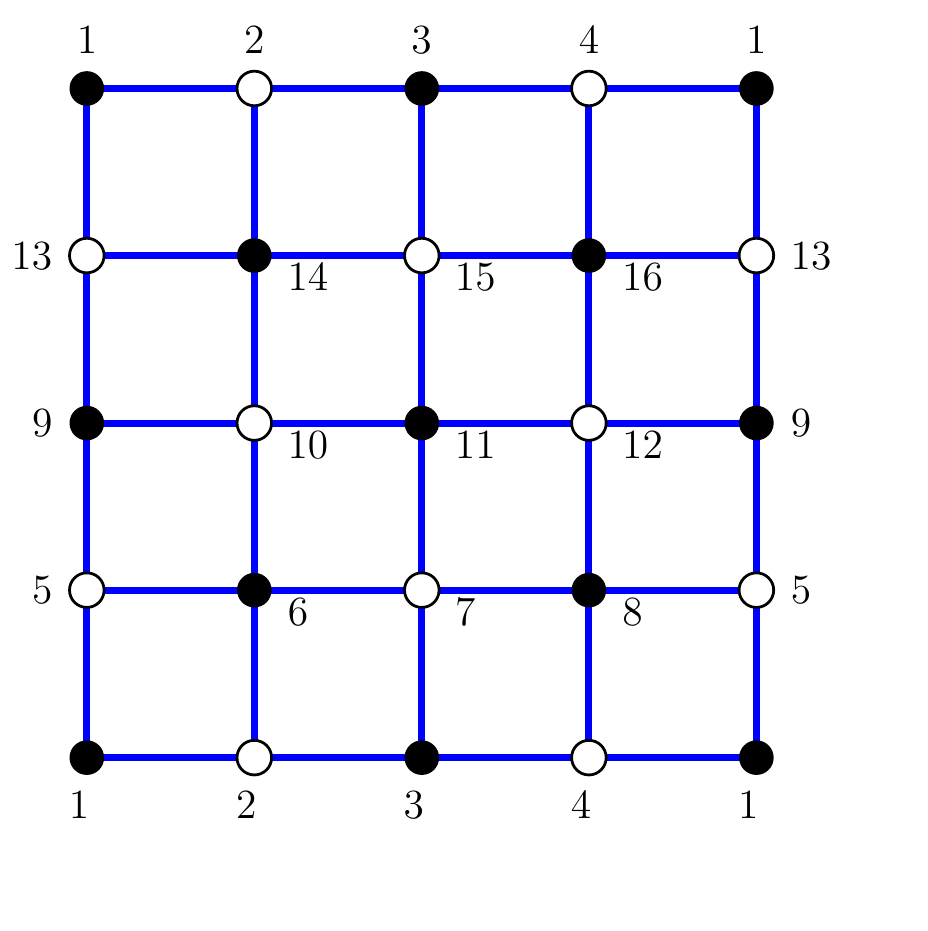}  \\[-7mm] 
  \hspace*{-6mm} (a) &\hspace*{-6mm} (b) \\ 
  \end{tabular}
  \vspace*{-2mm}
  \caption{%
  Constructing a quadrangulation of the class $\mathcal{Q}_0$.
  (a) The original graph is $G=(V,E) \simeq \mathrm{Sq}(4,2,2)$:
  the vertices are shown 
  with solid black dots, and the edges with thin black lines. 
  The dual graph $G^*=(V^*,E^*) \simeq G \simeq \mathrm{Sq}(4,2,2)$ is depicted
  with open black dots and thin dashed red lines. 
  Finally, we show the quadrangulation 
  $Q =(V\cup V^*, E_Q) \simeq \mathrm{Sq}(4,4)$:
  the edges in $E_Q$ are depicted with thick blue lines. We label the vertices 
  so that the boundary conditions in each graph are clear.  
  (b) We show the quadrangulation 
  $Q =(V\cup V^*, E_Q) \simeq \mathrm{Sq}(4,4) \in \mathcal{Q}_0$
  with the same notation as in (a). 
}
\label{fig_sq}
\end{figure}

Let us start with a connected graph $G=(V,E)$ embedded in a torus.
We can then consider (using the standard procedure) its
geometric dual graph $G^*=(V^*,E^*)$, which has a vertex in each face of $G$
and an edge crossing each edge of $G$; it
is also a connected graph on the torus.\footnote{
  We stress that the pair $(G,G^*)$ does {\em not}\/ in general possess
  all the usual properties for a dual pair of {\em planar}\/ graphs.
}
We then obtain a \emph{bipartite} quadrangulation 
$Q=(V_Q, E_Q)$ with vertex set $V_Q=V \cup V^*$ and edge set $E_Q$ that 
contain edges $\{i,j^*\}$ whenever $i\in V$ lies on the boundary of the 
face of $G$ that contains vertex $j^*\in V^*$. The graph $Q=Q(G)$ is indeed a 
bipartite quadrangulation: on each face of $Q$, one pair of 
diametrically opposite vertices corresponds to an edge $e\in E$, and the
other pair corresponds to the dual edge $e^* \in E^*$  
[see Fig.~\ref{fig_sq}(a)].
It is also easy to see
that every bipartite quadrangulation of the torus $Q$ arises via
this construction from some pair $(G, G^*)$.
Fig.~\ref{fig_sq} shows how the quadrangulation $Q(G)=\mathrm{Sq}(4,4)$ 
is obtained from $G=\mathrm{Sq}(4,2,2)$. 
 
%
%

\medskip

\noindent
{\bf Remarks.}
1. This procedure is a straightforward variant of the one outlined in 
Ref.~\cite{planar_AF_largeq} for quadrangulations of the plane 
(see also Refs.~\cite{selfdual1,selfdual2}).

2. Please note that there exist quadrangulations of the torus that 
are not bipartite, such as $\mathrm{Sq}(2p+1,2q+1)$ for $p,q\ge 1$;
these quadrangulations cannot be obtained via the above procedure. 
\myendremark

\medskip

At this point, it is useful to define
three classes of quadrangulations of the torus: 

\begin{definition} \label{def_Qs}
\hfill\break
\vspace*{-7mm}
\begin{itemize}
\item[a)] Let $\mathcal{Q}_0$ be the class of bipartite quadrangulations of 
          the torus.  
\item[b)] Let $\mathcal{Q}_1$ be the class of quadrangulations of the torus
          that have girth $\ge 4$.
\item[c)] Let $\mathcal{Q}$ be the class of all quadrangulations of the torus. 
\end{itemize}
\end{definition} 

\noindent
Indeed,
$\mathcal{Q}_0 \subset \mathcal{Q}_1 \subset \mathcal{Q}$,
 since every bipartite simple graph has girth $\ge 4$.
Furthermore, these inclusions are strict:
for instance, the quadrangulations $\text{Sq}(a,b)$
are non-bipartite whenever $a$ and/or $b$ is odd,
but they nevertheless have girth~4 whenever $a,b \ge 4$;
on the other hand, the quadrangulations $\text{Sq}(3,b)$
are non-bipartite and have girth~3.

Theorems~\ref{thm.bipartite} and~\ref{thm.main}(e$\,{}'$) imply:

\begin{corollary} \label{cor.quad}
\hfill\break
\vspace*{-7mm}
\begin{itemize}
 \item[(a)] WSK is ergodic for $q\ge 2$ on every (bipartite) quadrangulation 
            $Q \in \mathcal{Q}_0$. 
 \item[(b)] WSK is ergodic for $q\ge 4$ on every quadrangulation 
            $Q \in \mathcal{Q}_1$ (of girth $g\ge 4$).
\end{itemize}
\end{corollary} 

In particular, Corollary~\ref{cor.quad}(a)
applies to $G'_n(a,b)$ and $G''_n(a,b)$ whenever $a$ and $b$ are even,
and Corollary~\ref{cor.quad}(b) applies to these same graphs
whenever $a,b \ge 4$.

\bigskip

\noindent
{\bf Remarks.} 
1.  We do not know whether the condition ``girth $g \ge 4$''
in Corollary~\ref{cor.quad}(b) is really necessary.
We suspect that it is not.

2. In \cite{selfdual1,selfdual2} the following quadrangulations $Q=Q(G)$ were 
   considered. For self-dual $G$, the square, $Q(\text{hextri})$, 
   $Q(\text{house})$, $Q(\text{martini-B})$, and $Q(\text{cmm-pmm})$ lattices 
   \cite[Fig.~2]{selfdual2}.
   For non-self-dual $G$, the diced, $Q(\text{diced})$, $Q(\text{martini})$,
   $Q(\text{ruby})$, $Q(\text{cross})$, $Q(\text{asanoha})$, $G''_2$, and 
   $G''_3$ lattices \cite[Figs.~3, and~4]{selfdual2}. 
   If we choose their sizes appropriately, they belong to $\mathcal{Q}_0$; 
   otherwise, they belong to $\mathcal{Q}$.
\myendremark

\medskip

Conjecture~1.1 of Ref.~\cite{selfdual2} claims that those quadrangulations
$Q=Q(G)\in \mathcal{Q}_0$ with $G$ self-dual have
$q_c=3$ (i.e., the 3-state
AF Potts model on $G$ has a zero-temperature critical point)
and that, conversely, if $G$ is not self-dual, then $q_c > 3$.  

In particular, $G'_n$ and $G''_n$ belong to this second class. The behavior of
$q_c$ for $G'_n$ and $G''_n$ was studied in more detail in 
\cite{planar_AF_largeq}. In particular, $G'_n$ for $n\ge 2$, and $G''_n$ for 
$n\ge 4$ satisfy that $q_c > 4$. And $G'_n$ for $n\ge 4$, and $G''_n$ for 
$n\ge 6$ satisfy that $q_c > 5$. (Notice that $G''_4$ has $q_c=5.01(2)$, 
so this is a borderline case that deserves further investigations.) 
All of these cases are of physical relevance, as they imply the existence of 
phase transitions at nonzero temperature for $q=4$ and $q=5$, respectively.

The question is now to investigate the universality classes of these 
transitions. Ref.~\cite{planar_AF_largeq} explains that both $G_n$ 
and $H_n$ have the same transition as the underlying \emph{ferromagnetic} 
Potts model on the square lattice: i.e., second-order for $q\le 4$, and 
first-order for $q>4$. The ``extra'' edges in $G'_n$, $G''_n$, etc. are 
expected to be a ``small perturbation'', at least for large enough $q$. 
So one expects the behaviour of these modified lattices to be the same as 
$G_n$ and $H_n$; in particular, fist-order for $q>4$. One can conjecture 
that for large enough $q$, the transitions become first-order, as it happens 
for $H_n'''$ (see Section~\ref{sec.tri}). 

Finally, as $G'_n$ for $n\ge 2$, and $G''_n$ for $n\ge 4$, have $q_c>4$, 
they might have second-order transitions (at nonzero temperature) for $q=4$. 
These transitions could be of some interest.
  We do not know any example of a 4-state Potts antiferromagnet on a 
quadrangulation that undergoes a second-order transition,
and it seems to be an open question whether any exist.
  
In the same vein, as $G'_n$ for $n\ge 4$, and $G''_n$ for $n\ge 6$, have 
$q_c>5$, there might be a nonzero-temperature phase transition at $q=5$. 
Although one would expect a first-order phase transition in these cases, 
it also could possibly be a second-order transition.
This latter possibility could
be a lattice realization of the critical point found by Delfino and 
Tartaglia in the continuum \cite{Delfino_17}. If this second possibility 
becomes true, then the perturbation created by the ``extra edges'' in the
modified lattices should be large enough to change the nature of the transition.
 
%
%
\subsection{Triangulations of the torus} \label{sec.tri}

Let $Q = (V_Q,E_Q) \in\mathcal{Q}$ be a quadrangulation of the torus
(not necessarily bipartite).
Then we can construct a triangulation $T=(V_T,E_T)$
by adjoining a new vertex in each quadrangular face of $Q$, and four new edges
joining this new vertex with the four corners of the corresponding face. 
This new graph $T$ is an Eulerian triangulation with vertex set 
$V_T = V_Q \cup V_4$,
where $V_4$ is the set consisting of the ``new'' degree-4 vertices.
(We recall that a graph is called {\em Eulerian}\/
if every vertex has even degree. If a vertex $v \in V_Q$ had degree $d$ in $Q$,
then it has degree $2d$ in $T$.)
The edge set is $E_T=E_Q \cup E_4$ where $E_4$ is
the set consisting of the new edges.

\begin{figure}[htb]
\centering
  \begin{tabular}{cc}
  \includegraphics[width=0.47\columnwidth]{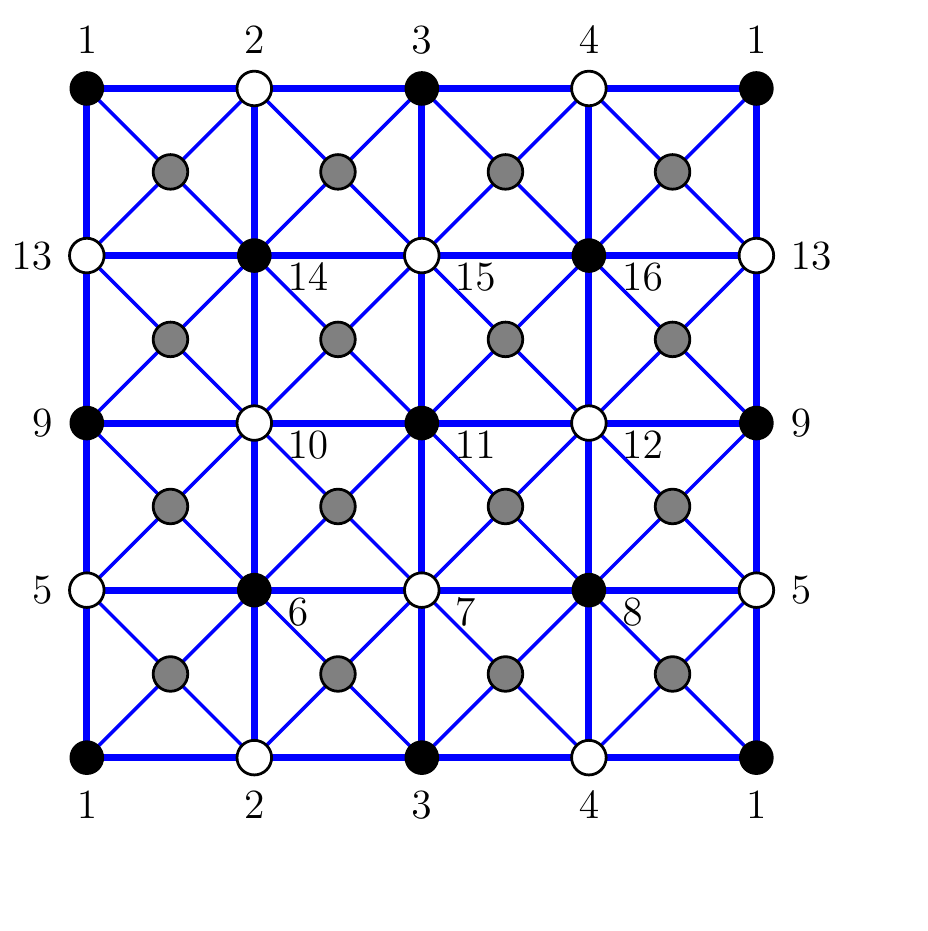} & 
  \includegraphics[width=0.55\columnwidth]{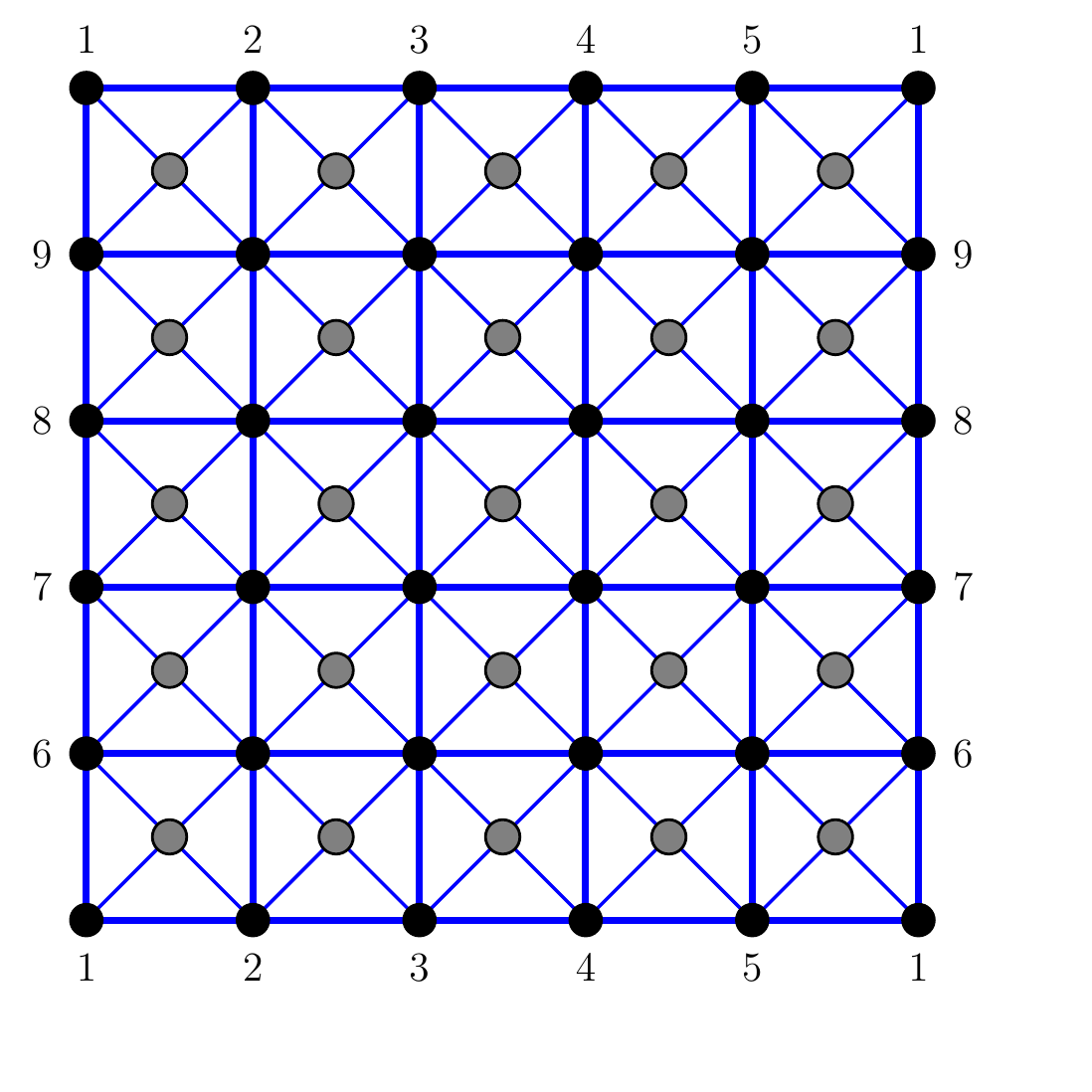} \\[-7mm] 
  \hspace*{-7mm} (a) &\hspace*{-6mm} (b) \\[-2mm] 
  \multicolumn{2}{c}{\includegraphics[width=0.34\columnwidth]{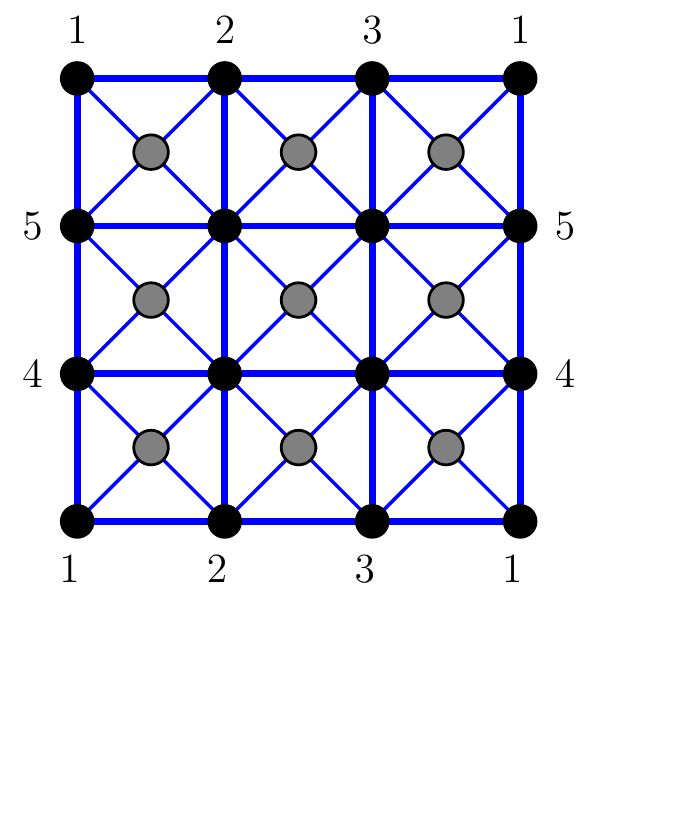}}
  \\[-18mm]
  \multicolumn{2}{c}{\hspace*{-6mm} (c)} 
  \end{tabular}
  \vspace*{-2mm}
  \caption{%
  Here $Q$ is the square lattice $\mathrm{Sq}(a,b)$,
  and $T$ is the union-jack lattice $\mathrm{UJ}(a,b)$.
  (a) $Q = \mathrm{Sq}(4,4) \in \mathcal{Q}_0$ [cf.~Fig.~\ref{fig_sq}(b)];
  the ``new'' degree-4 vertices are depicted in gray,
  and the ``new'' edges depicted as thin blue lines. The resulting 
  triangulation is $\mathrm{UJ}(4,4) \in \mathcal{T}_0$. 
  The labels mark the vertices of $\mathrm{Sq}(4,4)$.  
  (b) $Q = \mathrm{Sq}(5,5) \in \mathcal{Q}_1 \setminus \mathcal{Q}_0$,
  so that $T = \mathrm{UJ}(5,5) \in \mathcal{T}_1 \setminus \mathcal{T}_0$.
  Vertices with the same label should be indentified.
  (c) $Q = \mathrm{Sq}(3,3) \in \mathcal{Q} \setminus \mathcal{Q}_1$,
  so that $T = \mathrm{UJ}(3,3) \in \mathcal{T}_2 \setminus \mathcal{T}_1$.
  Vertices with the same label should be indentified.
}
\label{fig_tri}
\end{figure}

Suppose now that the quadrangulation $Q$ is bipartite
(i.e., belongs to $\mathcal{Q}_0$),
so that it arises from a graph $G = (V,E)$
and its geometric dual $G^* = (V^*,E^*)$ by the construction
of Section~\ref{sec.quad}.
Then the triangulation $T$ is tripartite:
one colour class (namely, $V_4$) contains degree-4 vertices,
while the other two ($V \cup V^*$)
induce the bipartite quadrangulation $Q$ [see Fig.~\ref{fig_tri}(a)]. 
Furthermore, in this case the triangulation $T$
is uniquely 3-colourable modulo permutations.
Conversely, every 3-colourable Eulerian triangulation 
of the torus in which one colour class
consists of degree-4 vertices and the 
other two induce a bipartite quadrangulation can be constructed in this way 
from some pair $(G,G^*)$. 

\medskip

{\bf Remark.}
This procedure is a straightforward variant
of the one outlined in Ref.~\cite{planar_AF_largeq}
for Eulerian triangulations of the plane. 
\myendremark

\medskip

\noindent
{\bf Examples.}
1. If $Q$ is the square lattice $\mathrm{Sq}(a,b)$,
then $T$ is the union-jack lattice $\mathrm{UJ}(a,b)$:
see Fig.~\ref{fig_tri}.
If $a$ and $b$ are even, then $Q$ is bipartite and $T$ is tripartite.
However, if $a$ and/or $b$ is odd,
then $Q$ has chromatic number~3 and $T$ has chromatic number~4.

2. If $Q$ is the diced lattice \cite{diced},
then $T$ is the bisected-hexagonal (BH) lattice (Fig.~\ref{fig_BH}).

3. One can obtain further examples of triangulations of this class by 
considering the bipartite quadrangulations of the torus shown in 
\cite{selfdual1,selfdual2} and following the above procedure. 
\myendremark 

Again it is useful to introduce some definitions:

\begin{definition} \label{def_Ts}
\hfill\break
\vspace*{-7mm}
\begin{itemize}
\item[a)] Let $\mathcal{T}_0$ be the class of Eulerian 
          triangulations of the torus such that their vertex set can be 
          partitioned as $V\cup V' \cup V_4$, so that the vertices in $V_4$ 
          have degree $4$, and the subgraph induced by $V\cup V'$ is a 
          bipartite quadrangulation of the torus
          (i.e., a member of $\mathcal{Q}_0$).

\item[b)] Let $\mathcal{T}_1$ be the class of Eulerian 
          triangulations of the torus such that their vertex set can be 
          partitioned as $V_Q \cup V_4$,
          so that the vertices in $V_4$ have degree $4$,
          and the subgraph induced by $V_Q$
          is a quadrangulation of the torus with girth $g\ge 4$
          {(i.e., a member of $\mathcal{Q}_1$).}

\item[c)] Let $\mathcal{T}_2$ be the class of Eulerian 
          triangulations of the torus such that their vertex set can be 
          partitioned as $V_Q \cup V_4$,
          so that the vertices in $V_4$ have degree $4$,
          and the subgraph induced by $V_Q$
          is a quadrangulation of the torus
          {(i.e., a member of $\mathcal{Q}$).}

\item[d)] Let $\mathcal{T}$ be the class of all triangulations of the torus. 
\end{itemize}
\end{definition}

%
%

Indeed, $\mathcal{T}_0 \subset \mathcal{T}_1 \subset \mathcal{T}_2 \subset
  \mathcal{T}$.
  Furthermore, as mentioned above,
  every $T \in \mathcal{T}_0$ is uniquely 3-colourable.
And of course, the
facial walk bounding every face of $T \in \mathcal{T}$ is a cycle of
length $g=3$.

The main result of this section is:

\begin{theorem} \label{theo.main.tri} 
\hfill\break
\vspace*{-7mm}
\begin{itemize}
   \item[(a)] WSK is ergodic for $q \ge 6$ on any triangulation 
              $T\in \mathcal{T}$.

   \item[(b)] WSK is ergodic for $q\ge 5$ on every triangulation  
              $T \in \mathcal{T}_1$.
\end{itemize}
\end{theorem}

\proof 
(a) This is a particular case of Theorem \ref{thm.main}(d${}'$). 

(b) The vertex set of $T$ has a partition $V_Q \cup V_4$
such that $V_4$ consists of degree-4 vertices,
and the subgraph induced by $V_Q$ is
a quadrangulation $Q\in\mathcal{Q}_1$.
Every $Q\in\mathcal{Q}_1$ is 4-degenerate by 
Corollary~\ref{cor.euler.2}(e),
so that WSK is ergodic on $Q$ for $q\ge 5$   
by Corollary~\ref{cor.euler.3}(e).
[Actually, WSK is ergodic on $Q$ for $q \ge 4$
   by Theorem~\ref{thm.main}(e${}'$), but we will not need this fact.]
If we now insert successively the degree-4 vertices of $V_4$,
Lemma~\ref{lemma.vergnas} proves the claim.
\qed 

\medskip

\noindent
{\bf Remarks.}
1. Theorem~\ref{theo.main.tri}(b) improves by one unit the 
   result of Theorem~\ref{theo.main.tri}(a), but for a smaller class of 
   triangulations of the torus.  

2. Theorem~\ref{theo.main.tri}(b) implies that WSK is ergodic 
   for $q\ge 5$ on all the triangulations $T\in \mathcal{T}_0$ obtained 
   from the quadrangulations studied in \cite{selfdual1,selfdual2}  
   (which include the UJ and the BH lattices). 
   In particular, 
   Theorem~\ref{theo.main.tri}(b) confirms the conjecture suggested in 
   Ref.~\cite{BH} for the BH lattice and $q=5$.

3. We do not know whether Theorem~\ref{theo.main.tri}(a)
   can be extended to $q=5$.

\begin{figure}[htb]
\centering
  \begin{tabular}{cc}
  \includegraphics[width=0.34\columnwidth]{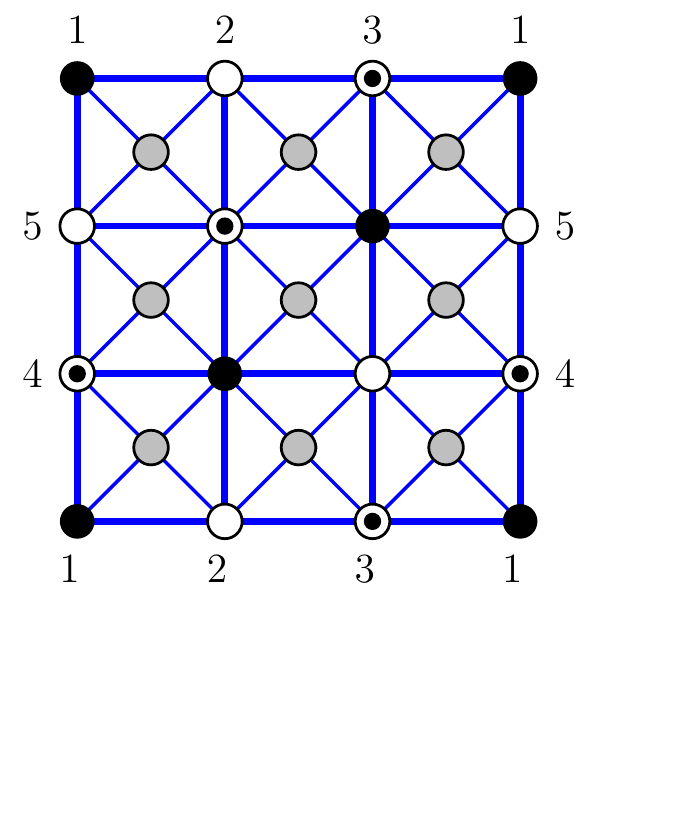} & 
  \includegraphics[width=0.34\columnwidth]{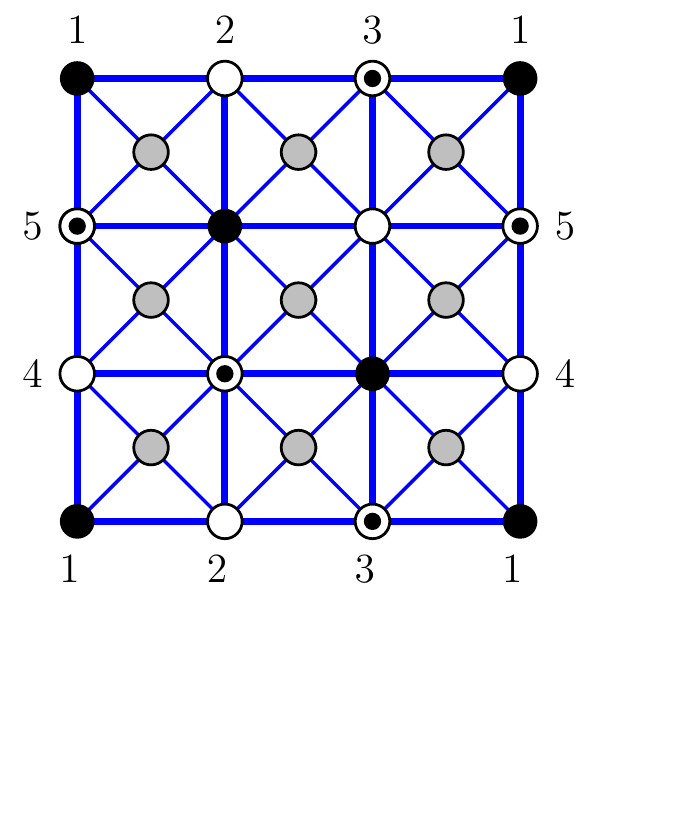} \\[-18mm] 
  \hspace*{-7mm} (a) &\hspace*{-7mm} (b) \\[2mm] 
  \end{tabular}
  \vspace*{-2mm}
  \caption{%
  The two 4-colourings (modulo global colour permutations) of the 
  $\text{UJ}(3,3)$ lattice. These two 4-colourings are not Kempe-equivalent.
  Vertices with the same label should be indentified.
}
\label{fig_UJ33}
\end{figure}

4. Theorem~\ref{theo.main.tri}(b), by contrast,
   \emph{cannot} be extended to $q=4$. 
   To see this, consider first the union-jack lattice $\mathrm{UJ}(3,3)$
   [Fig.~\ref{fig_tri}(c)],
   which has two 4-colourings modulo permutations,
   which are Kempe-inequivalent (see Fig.~\ref{fig_UJ33}).
   These two 4-colourings are obtained from the two Kempe-inequivalent
   3-colourings of the square lattice $\mathrm{Sq}(3,3)$ \cite{Lubin_Sokal}
   by colouring the degree-4 vertices of $\mathrm{UJ}(3,3)$
   with the fourth colour.
   These 3-colourings have constant colour
   either on SW--NE diagonals [Fig.~\ref{fig_UJ33}(a)]
   or on NW--SE diagonals [Fig.~\ref{fig_UJ33}(b)].
   Now, as observed in \cite{Lubin_Sokal},
   this same Kempe-inequivalence extends to the lattices $\mathrm{Sq}(3m,3n)$
   whenever $m,n$ are relatively prime, since each two-colour connected
   component then includes \emph{all} the vertices of those two colours.
   The same reasoning applies to the 4-colourings of $\mathrm{UJ}(3m,3n)$
   obtained by this construction from the 3-colourings of $\mathrm{Sq}(3m,3n)$,
   and shows that they are Kempe-inequivalent.
   The girth of the underlying quadrangulation $\text{Sq}(3m,3n)$ is
   $\min(4,3m,3n)$, which is $4$ whenever $m,n \ge 2$.

5. See, however, Theorem~\ref{theo.main.tri2} below for an extension of
   Theorem~\ref{theo.main.tri}(b) to $q=4$,
   for the \emph{smaller} class of triangulations $\mathcal{T}_0$.
\myendremark

%
%

The triangulations of greatest physical interest
are those with $q_c \ge 5$:
we wish to know whether some of them might have a
second-order phase transition (i.e., a critical point) for $q=5$.
One candidate was the BH lattice, but
it now appears \cite{BH} that the 5-state AF Potts model on the BH lattice
undergoes a very weak first-order phase transition at a finite temperature 
$v_c = -0.951\,308(2)$.

On the other hand, $H'''_2$ is an Eulerian triangulation with $q_c=5.26(2)$ 
\cite{planar_AF_largeq}. Again, the question is whether or not the 
``extra edges'' in $H'''_n$ for $n\ge 2$
are enough to change the nature of the transition at $q=5$ from first-order 
(as it is for $H_n$) to second-order.
Numerical results \cite{planar_AF_largeq} for $q\gtrsim 8$
show a first-order-transition behavior for $H_n'''$,
but for $4< q \lesssim 8$, the numerical data were inconclusive
(perhaps due to a large but finite correlation length).  
Again, a critical point at $q=5$ might be the confirmation of the results
by Delfino and Tartaglia \cite{Delfino_17}.

%
%
\section{Further results on triangulations of the torus} \label{sec.q=4}

In this section we are going to study further
the ergodicity of the WSK algorithm
for $q=4$ on triangulations belonging to the class $\mathcal{T}_0$. We will 
use methods pioneered by Fisk \cite{Fisk1,Fisk2,Fisk3} and also 
employed in \cite{MS1,MS2}. These methods (based on algebraic topology) can 
only be applied to 4-colourings. In this section, 4-colourings that are related
by a global permutation of colours are identified. 

%
%
\begin{figure}[tbh]
\centering
  \includegraphics[width=0.2\columnwidth]{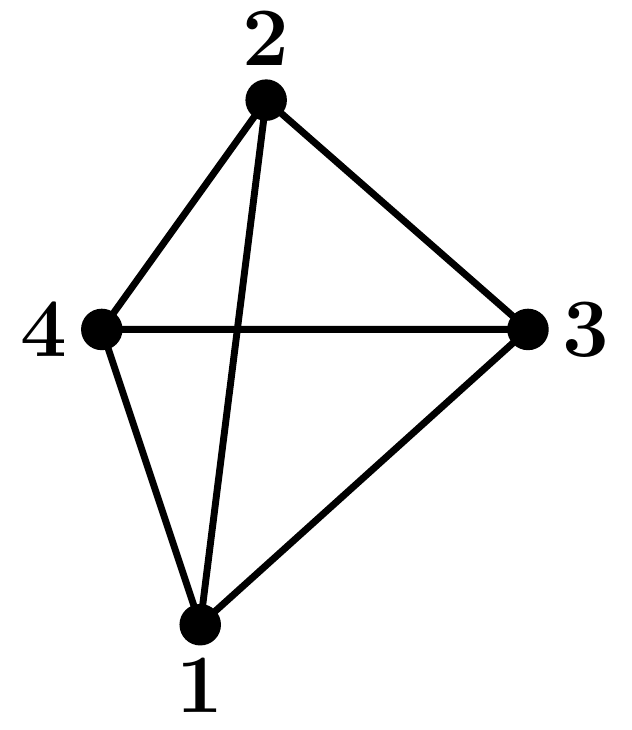}  
  \vspace*{-2mm}
  \caption{%
  The tetrahedron $\Delta^3$. Its surface $\partial \Delta^3$ can be 
  considered a triangulation of the sphere $S_0$.  
}
\label{fig_tetra}
\end{figure}

We recall that $S_p$ denotes the (orientable) surface obtained from the sphere
by attaching $p$ handles ($p \ge 0$);
in particular, $S_1$ is the torus.
Let $T \subset S_p$ be a triangulation
that is cellularly embedded in $S_p$.
In Fisk's approach, a proper 4-colouring $f$ of $T$ is considered to be
a non-degenerate simplicial map \cite{Fisk1,Fisk2}
\be
   f \colon\: S_p \to \partial \Delta^3  
\ee
where $\partial \Delta^3$ is the surface of a tetrahedron, which can 
be considered a triangulation of the sphere $S_0$ (see Fig.~\ref{fig_tetra}). 
A non-degenerate simplicial map $f \colon S_p \to \partial \Delta^3$
is a map such that the image of every triangle of $T$ under $f$
is a triangle of $\partial \Delta^3$. 

From algebraic topology \cite{Fisk2},
if $T$ is a triangulation of an orientable closed surface $S_p$
and $f \colon S_p \to \partial \Delta^3$ is a non-degenerate simplicial map,
then there exists an integer-valued quantity
$\text{deg}(f)$, called the \emph{degree} of $f$,
that is determined up to a sign.
The degree of a 4-colouring can be computed
as follows. First, one should choose orientations for both
$S_p$ and $\partial \Delta^3$. Then, given any triangle $t$ of 
$\partial \Delta^3$ (say, $t=123$), 
which corresponds to a particular three-colouring of a triangular face, we
can compute the number $p$ (resp.~$r$) of triangles of $T$ mapping to $t$
which have their orientation preserved (resp.~reversed) by $f$. Then, the
degree of the 4-colouring $f$ is\footnote{
   See \cite[p.~327]{Fisk1} and \cite[pp.~303--304]{Fisk2}.
   See also \cite[p.~51]{Croom_78} and \cite[pp.~24--25]{Outerelo_09}.
}
\be
\text{deg}(f) \;=\; p - r \,.
   \label{eq.degf}
\ee 
In particular, the combination~\reff{eq.degf}
does not depend on the chosen triangle $t$ of $\partial \Delta^3$.
Note also that if $f$ is actually a 3-colouring,
then its degree is zero:
for instance, if $f$ uses only the colours 1, 2 and 3, then
we can choose $t=124$, so that $p=r=0$.
In practice we will consider only $|\text{deg}(f)|$, as two 
4-colourings related by a global colour permutation will be
identified. 

The key result in this section is due to Fisk \cite[Theorem~37]{Fisk3}:

\begin{theorem} \label{thm.fisk}
Let $T$ be a triangulation of the sphere or the torus.
If $T$ is 3-colourable,
then all 4-colourings with degree divisible by $12$ are Kempe equivalent.
\end{theorem}

Using this theorem, we can prove the following:

\begin{theorem} \label{theo.main.tri2}
WSK is ergodic for $q\ge 3$ on every triangulation $T \in \mathcal{T}_0$.
\end{theorem}

\begin{figure}[htb]
\centering
  \includegraphics[width=0.99\columnwidth]{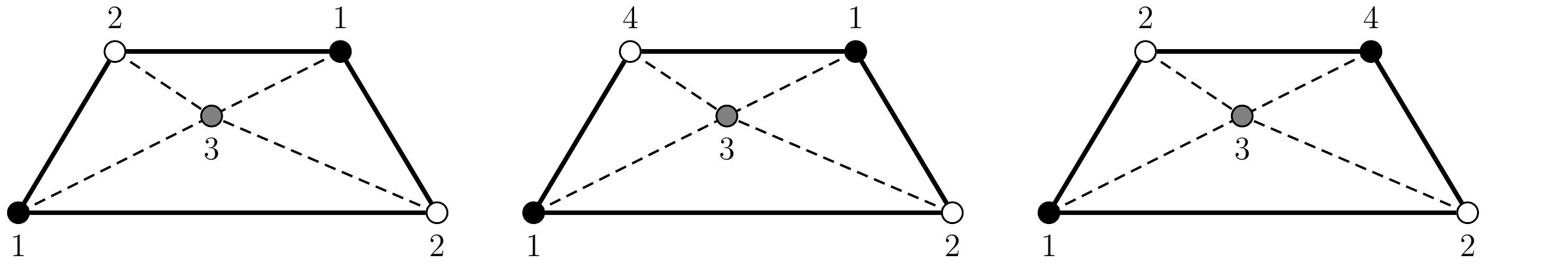}  
  \vspace*{-2mm}
  \caption{%
  The three inequivalent colourings of a quadrangle of $T\in\mathcal{T}_0$. 
  The black (resp.\ white) dots belong to $V$ (resp.\ $V'$), while the 
  gray dot belongs to $V_4$. The solid lines are edges that belong to both
  the quadrangulation and the triangulation, while the dashed lines represent
  edges belonging only to the triangulation. 
}
\label{fig_Q}
\end{figure}

\proof 
The result is trivial for $q=3$,
because every triangulation $T \in \mathcal{T}_0$
is uniquely 3-colourable modulo permutations.
The case $q=5$ is covered by
Theorem~\ref{theo.main.tri}(b),
since
$\mathcal{T}_0 \subset \mathcal{T}_1$.
So it suffices to consider $q=4$.
Let us consider a triangulation $T\in \mathcal{T}_0$ of the torus. By 
construction it is Eulerian and 3-colourable. This triangulation has vertex
set $V \cup V' \cup V_4$, so that $V_4$ consists in degree-4 vertices, and 
$V\cup V'$ induce a bipartite quadrangulation of the torus 
$Q\in \mathcal{Q}_0$. If we look at each quadrangle of $Q$, we see the
structure shown in Fig.~\ref{fig_Q}. 

%
%

\noindent
There are only three distinct 4-colourings of each quadrangle (modulo global 
permutations of colours). The important observation is, no matter what
triangle $t$ of $\partial \Delta^3$ we choose, the net contribution of each
colouring to the degree is always zero. As all these quadrangles form the 
triangulation $T$, the degree of any 4-colouring $f$ of $T$ will have 
$\text{deg}(f)=0$. Therefore, using Theorem~\ref{thm.fisk}, we conclude
that all 4-colourings of $T$ are Kempe 4-equivalent. In other words, 
WSK is ergodic on $T$ for $q=4$. \qed

\medskip

Along the way, we have also proven:

\begin{proposition}   \label{prop.deg0}
Every 4-coloring of a triangulation $T \in \mathcal{T}_0$ has degree~0.
\end{proposition}

This can be compared with \cite[Theorem~1]{Fisk1},
which proves the same result for Eulerian triangulations of the sphere.

\bigskip

\noindent
{\bf Remarks.} 1.
Theorem~\ref{theo.main.tri2} can be contrasted with the result of \cite{MS1}:
   WSK for $q=4$ is not ergodic on any (6-regular) triangular lattice of 
   size $3m\times 3n$ on the torus with $m,n\ge 2$.
It can also be contrasted with the fact, remarked in Section~\ref{sec.tri},
that WSK for $q=4$ is not ergodic on any union-jack lattice
$\mathrm{UJ}(3m,3n)$ on the torus where $m,n$ are relatively prime.

2. Theorem~\ref{theo.main.tri2} improves the result of 
   Theorem~\ref{theo.main.tri}(b) by one unit, but the class of triangulations 
   is smaller.
\myendremark

%
%
\section{Summary}  \label{sec.summary}

In this paper we have studied the ergodicity at zero temperature
of the WSK algorithm for the $q$-state Potts antiferromagnet
on various classes of graphs embedded on the torus.
Using graph-theoretic methods (Section~\ref{sec.torus}),
we have shown that
the WSK algorithm is ergodic for $q\ge 4$ on any quadrangulation
of the torus of girth $\ge 4$ (what we have called class $\mathcal{Q}_1$),
and that it is ergodic for $q \ge 5$
on any Eulerian triangulation of the torus such that one sublattice
consists of degree-4 vertices while the other two sublattices
induce a quadrangulation of girth $\ge 4$
(what we have called class $\mathcal{T}_1$).
Furthermore, using methods from algebraic topology pioneered by Fisk
(Section~\ref{sec.q=4}),
we have shown that the WSK algorithm is ergodic for $q \ge 3$
on any Eulerian triangulation of the torus such that one sublattice
consists of degree-4 vertices while the other two sublattices
induce a bipartite quadrangulation
(what we have called class $\mathcal{T}_0$,
 which is properly contained in $\mathcal{T}_1$).
These classes include many lattices of interest in statistical mechanics
(Section~\ref{sec.application}).

Finally, we have shown by explicit counterexamples
that many of our results are sharp.
However, in at least two cases we do not know whether this is so,
and we have posed these as open problems
(Questions~\ref{question.trifree.q=3} and \ref{question.toroidal.q=5}).

%
%
\section*{Acknowledgments}

We warmly thank Jesper Jacobsen and Bojan Mohar for a careful reading of 
early drafts of the manuscript, and correspondence. 

The authors' research was supported in part 
by the Spanish Ministerio de Econom\'{\i}a, Industria y Competitividad 
(MINECO), Agencia Estatal de Investigaci\'on (AEI), and Fondo Europeo de
Desarrollo Regional (FEDER) through grant No. FIS2017-84440-C2-2-P, 
by grant No.~PID2020-116567GB-C22 AEI/10.13039/501100011033, 
by the Madrid Government (Comunidad de 
Madrid-Spain) under the Multiannual Agreement with UC3M in the line of 
Excellence of University Professors (EPUC3M23), and in the context of the 
V~PRICIT (Regional Programme of Research and Technological Innovation), and 
by U.K.~Engineering and Physical Sciences Research Council
grant EP/N025636/1.

%
%

\end{document}